\documentclass[12pt]{article}

\usepackage{latexsym,amsmath,graphicx,psfrag,ifthen}    

\renewcommand{\theequation}{\arabic{section}.\arabic{equation}}

\newcommand{\dega}{\dagger}

\newcommand{\br}{\langle}
\newcommand{\ke}{\rangle}

\newcommand{\sss}{\scriptscriptstyle}

\newcommand{\bda}{\begin{\displaymath}\begin{array}{rl}}
\newcommand{\eda}{\end{array}\end{displaymath}}
\newcommand{\be}{\begin{equation}}
\newcommand{\ee}{\end{equation}}
\newcommand{\bdm}{\begin{displaymath}}
\newcommand{\edm}{\end{displaymath}}
\newcommand{\bea}{\begin{eqnarray}}
\newcommand{\eea}{\end{eqnarray}}
\newcommand{\no}{\nonumber \\}

\newcommand{\fs}{\; .}
\newcommand{\co}{\; ,}
\newcommand{\al}{&\!\!}
\newcommand{\eff}{{e\hspace{-0.1em}f\hspace{-0.18em}f}}

\newcommand{\QCD}{\mbox{\tiny Q\hspace{-0.05em}CD}}
\newcommand{\indV}{\mbox{\tiny V}}

\newcommand{\indL}{{\scriptscriptstyle L}}
\newcommand{\indR}{{\scriptscriptstyle R}}
\newcommand{\lvac}{\langle 0|\,}
\newcommand{\rvac}{\,|0\rangle}
\newcommand{\wave}{\raisebox{0.22em}{\fbox{\rule[0.15em]{0em}{0em}\,}}\,}

\newcommand{\nc}{N_{\!c}}
\newcommand{\nf}{N_{\!f}}

\newcommand{\qbar}{\overline{\rule[0.42em]{0.4em}{0em}}\hspace{-0.5em}q}
\newcommand{\ubar}{\overline{\rule[0.42em]{0.4em}{0em}}\hspace{-0.5em}u}

\newcommand{\Ubar}{\bar{U}}
\newcommand{\LSU}{L^{\mbox{\tiny SU}_{\hspace{-0.07em}3}}}
\newcommand{\HSU}{H^{\mbox{\tiny SU}_{\hspace{-0.07em}3}}}
\newcommand{\SU}{\mbox{\tiny SU}_{\hspace{-0.07em}3}}

\newcommand{\spc}{\hspace{2em}}

\newcommand{\za}{Z_{\!\sss{A}}}
\newcommand{\zm}{Z_{\,\!\sss{\bar{q} q}}}
\newcommand{\ga}{\gamma_{\!\sss{A}}}
\newcommand{\gm}{\gamma_{\,\!\sss{\bar{q} q}}}

\newcommand{\non}{\nonumber}

\newcommand{\tpsi}{\bar{\psi}}
\newcommand{\indtheta}{\mbox{\raisebox{-0.15em}{$\scriptstyle \theta$}}}
\newcommand{\qm}{m}
\newcommand{\RL}{(\mbox{\it \small R}\leftrightarrow \mbox{\it\small L})}
\newcommand{\tN}{\vartheta}

\newcommand{\ind}{\scriptscriptstyle}
\newcommand{\chitheta}{\mbox{\raisebox{0.15em}{$\chi\rule[-0.4em]{0em}
{0em}_\theta$}}}
\def\boldm#1{{\boldmath #1 \unboldmath}}

\begin{document}

\begin{titlepage}
\begin{flushright}BUTP-00/19\end{flushright}
\vspace{2.5cm}
\begin{center}
{\LARGE\bf Large \boldm{$\nc\!\!$} in chiral perturbation theory}

\vspace{0.5cm}

\vspace{0.5cm}
R.~Kaiser and H.~Leutwyler\\Institute for Theoretical Physics,
University of Bern,\\  
Sidlerstr. 5, CH--3012 Bern, Switzerland\\
E-mail: kaiser@itp.unibe.ch, leutwyler@itp.unibe.ch

\vspace{0.5cm}
July 2000

\vspace{1cm}

\begin{abstract}
The construction of the effective
Lagrangian relevant for the mesonic sector of QCD in the large $\nc$ limit
meets with a few rather subtle problems. We thoroughly examine these and 
show that, if the variables of the effective theory are chosen suitably, 
the known large $\nc$ counting
rules of QCD can unambiguously be translated into corresponding
counting rules for the effective coupling constants. As an application,
we demonstrate that the Kaplan-Manohar transformation is in conflict with
these rules and is suppressed to all orders in $1/\nc$.
The anomalous dimension of the axial singlet current generates 
an additional complication: The corresponding external field undergoes
nonmultiplicative renormalization. As a
consequence, the Wess-Zumino-Witten term, which accounts 
for the U(3)$_{\indR}\times$U(3)$_{\indL}$ anomalies in the framework of the
effective theory, contains pieces that depend on the running scale of QCD. 
The effect only shows up at nonleading order in
$1/\nc$, but requires specific unnatural parity contributions 
in the effective Lagrangian that restore renormalization group invariance.
\end{abstract}

\vspace{1cm}
\footnotesize{\begin{tabular}{ll}
{\bf{Pacs:}}$\!\!\!\!$& 11.15.Pg, 11.30.Rd, 12.38.Aw, 12.39.Fe, 14.40.-n\\
{\bf{Keywords:}}$\!\!\!\!$& QCD, $1/\nc$ expansion,  
effective Lagrangian,   
chiral symmetry,\\&  anomalies, chiral perturbation
theory
\end{tabular}}
\vspace{1.5cm}

\rule{33em}{0.02em}\\
{\footnotesize Work supported in part by Schweizerischer Nationalfonds}  
\end{center}
\end{titlepage}

\tableofcontents
\clearpage
\setcounter{equation}{0}
\section{Introduction}
\label{intro}
The low energy properties of QCD are governed by an approximate,
spontaneously broken symmetry, which originates in the fact that three of the
quarks happen to be light. If $m_u,m_d,m_s$ are turned off, the symmetry
becomes exact. The spectrum of the theory then contains eight
massless
pseudoscalar mesons, the Goldstone bosons connected with the spontaneous
symmetry breakdown.

If the number of colours is taken large, the quark
loop graph which gives rise to the U(1)-anomaly 
is suppressed \cite{Large Nc}. This implies that, in the limit
$\nc\rightarrow\infty$, the singlet axial current is also conserved: The 
theory in effect acquires a higher degree of symmetry. Since the operator
$\qbar q$ fails to be invariant under the extra U(1)-symmetry, the formation
of a quark
condensate, $\lvac \qbar q\rvac\neq 0$, implies that this symmetry is also
spontaneously broken \cite{Coleman Witten}. The spectrum of QCD, therefore, 
contains a ninth
state, the $\eta^\prime$, which
becomes massless if not only $m_u,m_d,m_s$ are turned off, but if in addition
the number of colours is sent to infinity. 

Chiral symmetry imposes strong constraints on the
properties of the Goldstone bosons. These may be worked out in a
systematic manner by means of 
the effective Lagrangian method, which describes the low energy
structure of the theory in terms of an expansion in powers of
energies, momenta and quark masses \cite{Weinberg Physica,GL SU(3)}.
The fact that, in
the large $\nc$ limit, the $\eta'$ also plays the role of a Goldstone boson 
implies that the properties of this particle are subject to analogous
constraints, which may again be worked out by means of a suitable
effective Lagrangian. 

The main features of the effective theory relevant for the meson 
sector of QCD in the large $\nc$ limit were discovered long ago \cite{Leff
  U(3)}. The systematic analysis in the framework of chiral perturbation
theory was taken up in ref.~\cite{GL SU(3)}, where the Green functions 
of QCD were studied by means of a 
simultaneous expansion in powers of momenta, quark masses and $1/\nc$.
A considerable amount of work concerning the structure of QCD 
at large $\nc$ has been carried out since then 
\cite{Christos}--\cite{Herrera 4} and quite a few
phenomenological applications   
of the $1/\nc$ expansion
have appeared in the literature. For a review of these, in particular also 
for a discussion
of the $\eta$-$\eta'$ mixing pattern, we refer
to \cite{Feldmann}. Moreover, the large $\nc$ properties 
of the effective theory relevant for the baryons were recently investigated in
detail \cite{Baryons}.
  
In the present paper, we examine the foundations of the effective theory in 
the sector with baryon number $B=0$ and show that this leads to new 
insights into the low energy structure of QCD in the large $\nc$ limit.
The framework is more
complicated than in the case of $\nc=3$, because an additional  
low energy scale appears, related to the
mass of the $\eta'$.  In order to firmly establish our claims,
we first need to put the effective theory on a solid basis and demonstrate
that the effective coupling constants necessarily obey
the counting rules indicated in ref.~\cite{GL SU(3)}
(sections \ref{effective action}--\ref{dep theta} and appendix 
\ref{appendix A}). Using these, we then show  that the transformation
introduced by Kaplan and  
Manohar is in conflict with the large $\nc$ properties of QCD: The
parameter $\lambda$ occurring in the transformation
\bea\label{KMintro} m_u\rightarrow m_u+\lambda
\,m_d\,m_s\hspace{2em}(\mbox{cyclic} \;\;
u\rightarrow d\rightarrow s\rightarrow u)
\co\eea
vanishes to all orders of the $1/\nc$ expansion
(section \ref{KaplanManohar}). Next, we discuss the consequences of the fact 
that the
dimension of the singlet axial current is anomalous and determine the 
corresponding dependence of the effective coupling constants
on the running scale of QCD (sections \ref{anomalous}--\ref{coupling}). 
The matching between
the effective theories relevant for a finite and an infinite number of
colours is worked out in section \ref{comparison} and appendix
\ref{matching2}. Finally, in sections  
\ref{unnatural}--\ref{unnatural ho} and appendix \ref{decomposition}, 
we examine the modifications required to extend the Wess-Zumino-Witten term
from SU(3)$_{\indR}\times$SU(3)$_{\indL}$ to 
U(3)$_{\indR}\times$U(3)$_{\indL}$.

Some of the results described in the present paper were reported earlier
\cite{Adelaide}. The application to the masses, decay constants and photonic
transitions of the pseudoscalar mesons will be given elsewhere \cite{KL in
  preparation}. 
A further application concerns the low energy properties of the correlation
function
\bea\label{eq:omega} 
\chi(q^2)=-i\!\!\int dx\, e^{iq\cdot x}\lvac T\omega(x)\omega(0)\rvac\co
\hspace{2em} \omega=\frac{1}{16\pi^2}\,\mbox{tr}
\hspace{-0.5em}\rule[-0.5em]{0em}{0em}_c\hspace{0.4em}
G_{\mu\nu}\tilde{G}^{\mu\nu}
\co\eea
where we have absorbed the coupling constant in the gluon field.
In particular, the manner in which the topological susceptibility 
$\chi(0)$ and the derivative
$\chi'(0)$ depend on the light quark masses is quite remarkable \cite{KL in
  preparation, topsusz}. 

\setcounter{equation}{0}
\section{Effective action}\label{effective action}
Our analysis is based on the known large $\nc$ behaviour of QCD. 
We work with the effective action of this theory, which describes the response
of the system to the perturbation generated by a set of external fields,
\bea {\cal L}_{\QCD}={\cal L}^0_{\QCD} +\qbar\gamma^\mu(v_\mu+\gamma_5 a_\mu)
q- \qbar(s-i\gamma_5 p)q -\theta\, \omega\fs\eea
The term ${\cal L}^0_{\QCD}$ describes the limit where the
masses of the three light quarks and the vacuum angle are set to 
zero. The external fields $v_\mu(x)$, $a_\mu(x)$, $s(x)$, $p(x)$ represent
hermitean $3\times 3$
matrices in flavour space. The mass matrix of the three light quarks
is contained in the scalar external field
$s(x)$. The vacuum angle $\theta(x)$ represents the variable conjugate
to the operator $\omega(x)$ specified in eq.~(\ref{eq:omega}).
In Euclidean space, the integral
\bea \nu=\int\!\! dx\,\omega \nonumber\eea
is the winding number of the gluon field, so that $\omega(x)$ may be
viewed as the winding number density. 

The effective action represents the logarithm of the vacuum-to-vacuum
transition amplitude,
\bea e^{iS_\eff\{v,a,s,p,\theta\}}=\langle 0\,\mbox{out}|
0\,\mbox{in}\rangle\rule[-0.3em]{0em}{0em}_{v,a,s,p,\theta}\fs\eea
The coefficients of the expansion of
$S_\eff\{v,a,s,p,\theta\}$ 
in the external fields are the connected correlation functions 
of the vector, axial, scalar and pseudoscalar
quark currents and of the winding number density, in the massless theory.
Lorentz invariance implies that these can be decomposed into scalar functions,
with coefficients that contain the external momenta and the tensors
$g_{\mu\nu}$, $\epsilon_{\mu\nu\rho\sigma}$. In view of the fact that the
square of $\epsilon_{\mu\nu\rho\sigma}$ can be expressed in terms of 
$g_{\mu\nu}$, there are 
two categories of contributions: The natural parity part of the effective
action, which collects the pieces that do not contain the $\epsilon$-tensor, 
and the unnatural parity part, where this
tensor occurs exactly once. 

The consequences of the symmetry properties of QCD for the natural parity part
of the effective action are remarkably simple: 
The Ward identities related to the conservation of the
vector and axial currents imply that this
part of the effective action is invariant under chiral 
gauge transformations,
\bea \label{trafo}
\al\al r_\mu'= V_{\indR} r_\mu V_{\indR}^\dagger + i V_{\indR} \partial_\mu
V_{\indR}^\dagger\co\hspace{2em} 
l_\mu'= V_{\indL} l_\mu V_{\indL}^\dagger + i V_{\indL} \partial_\mu
V_{\indL}^\dagger\\
\al\al s'+ip'= V_{\indR}(s+ip)V_{\indL}^\dagger\co\hspace{2.4em}  
\theta'= \theta +i\ln\det V_{\indR}-i\ln\det V_{\indL} \co\nonumber\eea
with $r_\mu=v_\mu+ a_\mu$, $l_\mu=v_\mu-a_\mu$ and
$V_{\indR}(x), 
V_{\indL}(x)\in\mbox{U(3)}$. 

Chiral U(1) transformations play an 
essential role for the properties of the
theory at large $\nc$. It is important that our list of external fields
includes a source for the singlet axial current: the trace of the
matrix $a_\mu$. The divergence of this current contains
an anomaly proportional to $\omega$. The above transformation law
automatically accounts for this term, through the shift
in the vacuum angle that is generated by a U(1) rotation.

The U(1) anomaly is not the only one relevant in our context, 
but the remaining anomalies of the effective action are independent of the 
interaction, so that they only affect the unnatural parity part. 
We first investigate the low energy structure
of the natural parity part, which, as stated above, is gauge invariant:
$S_\eff\{v',a',s',p',\theta'\}_{\mbox{\tiny nP}}=
S_\eff\{v,a,s,p,\theta\}_{\mbox{\tiny nP}}$. 
The unnatural parity part -- in particular, the Wess-Zumino-Witten term --
only contributes at nonleading orders of the low energy expansion and 
will be discussed in detail later on. 

\setcounter{equation}{0}
\section{QCD at large \boldm{$\nc$}}
\label{gluodynamics}

The well-known leading logarithmic formula
\bea \frac{g^2}{(4\pi)^2}=\frac{1}{\beta_0 \ln(
  {\mu^2}/{\Lambda_{\QCD}^2})}
\co\hspace{2em}
\beta_0=\frac{1}{3}\,(11\nc-2 \nf)\eea
implies that the running coupling constant tends to zero when $\nc$
becomes
large, $g^2\sim 1/\nc$. At leading order of the $1/\nc$ expansion, the
Green 
functions are dominated by those graphs that contain the smallest
possible
number of quark loops. For the correlation functions of the
quark currents, graphs with one such loop generate the leading contributions. 

Consider the connected correlation function 
formed with $n_j$ quark currents $j_i=\qbar\,\Gamma_i q$ and $n_\omega$ winding
number densities
\bea
G_{n_j n_\omega}=\lvac Tj_1(x_1)\cdots j_{n_j}(x_{n_j})\,\omega(y_1)
\cdots \omega(y_{n_\omega})\rvac\rule[-0.2em]{0em}{0em}_c\eea
and denote the fraction due to graphs with $\ell$ quark loops by
$G_{n_j n_\omega}^{\,\ell}$. The large $\nc$ counting rules of perturbation 
theory
imply that this quantity represents a term of order 
\bea \label{countG}
G_{n_j  n_\omega}^{\,\ell}=O(\nc^{2-\ell-n_\omega})
\co\hspace{2em}\ell=0,1,\ldots\eea  
The leading power is 
independent of the
number $n_j$ of quark currents, but decreases with $n_\omega$. 
This implies that 
the dependence of the effective action on the vacuum angle is
suppressed. The leading
terms in the $1/\nc$ expansion of the effective action may
be characterized by the formula
\bea S_\eff=\nc^2\,S_0\{\tN\}+\nc\,
S_1\{v,a,s,p,\tN\}+
S_2\{v,a,s,p,\tN\}+\ldots
\co\label{G2}\eea
where $S_\ell$ collects the contributions from graphs with $\ell$ quark loops
and 
\bea\tN\equiv\frac{\theta}{\nc}\fs\eea
Since the external fields $v,a,s,p$ contained in $S_1$ are attached 
to one and the same quark
loop, the expansion of $S_1$ in terms of these fields 
generates expressions that 
involve a single trace over the flavour indices. Likewise, $S_2$ represents 
a sum of contributions containing at most two flavour traces, etc.

The dominating contributions to the effective action
arise from those graphs that do not involve quark lines.  Their sum represents
the effective action of gluodynamics, which only depends on the vacuum
angle. The expansion in powers of $\theta$ yields the connected correlation
functions of the field $\omega(x)$, \bea
S^{\mbox{\tiny{GD}}}_{\eff}\al=\al\frac{i}{2!}\int \!\!dx_1\,
dx_2\,\theta(x_1) \,\theta(x_2)\langle
0|T\omega(x_1)\,\omega(x_2)|0\rangle_{\mbox{\tiny GD}} \no &&
-\frac{i}{4!}\int\!\!dx_1\ldots dx_4\, \theta(x_1)\cdots\theta(x_4)\langle
0|T\omega(x_1)\cdots\omega(x_4)|0\rangle_{\mbox{\tiny GD}}+
\ldots\nonumber\eea 
The leading contribution in the $1/\nc$ expansion of the effective action
of gluodynamics coincides with the one occurring in QCD,
\bea S^{\mbox{\tiny{GD}}}_{\eff} =\nc^2
S_0\{\tN\}+O(\nc)\fs\eea 
In particular, the topological susceptibility
of gluodynamics\footnote{As written, the formula 
refers to Euclidean space.
In Minkowski space, the expression reads 
\bea \tau_{\mbox{\tiny GD}} =-i\!\!\int\!\! dx\lvac T \omega(x)\, 
\omega(0)\rvac_{\mbox{\tiny GD}}\fs\nonumber\eea},
\bea \tau_{\mbox{\tiny GD}} \al\equiv\al\frac{\langle
  \nu^{\,2}\rangle}{V}\hspace{-0.05em} \mbox{\raisebox{0.4em}{\tiny GD}}
=\int\!\! dx\lvac T \omega(x)\, \omega(0)\rvac_{\mbox{\tiny GD}}\nonumber\co
\eea
represents a term of order 1.

Since gluodynamics possesses a mass gap (excitation energy of
the lightest glueball), the correlation functions of $\omega$ decay
on a scale of order $\Lambda_{\QCD}$. We are analyzing the theory for
external fields that vary slowly on this scale. 
In that regime, we may expand the factor $\theta(x_1)\ldots\theta(x_n)$
around the point $x_1$ and express the functional through a 
derivative expansion:
\bea\label{derivative expansion} S_0\{\tN\}=\int\!\!dx\left\{-e_0(\tN)
+\partial_\mu\tN\,\partial^\mu\tN\, 
e_1(\tN)+
\ldots\right\}\co\label{S0}\eea
where $e_0(\tN), e_1(\tN),\ldots$ are ordinary functions of a single variable.
The quantity $\nc^2e_0(\tN)$ 
represents the energy density of the vacuum for the case where,
in the vicinity of the point under consideration, the vacuum angle is taken
constant. The term
$\nc^2\partial_\mu\tN\,\partial^\mu\tN\, e_1(\tN)$ describes the 
change in the energy density that arises if the vacuum angle has a
nonvanishing gradient there, etc.

\setcounter{equation}{0}
\section{Relation between QCD and gluodynamics}\label{relation}
The large $\nc$ counting rules imply that the quantities 
$e_0(\tN)$, $e_1(\tN)$, $\ldots$ , which occur in the derivative expansion
(\ref{derivative expansion}) of the  effective action of 
gluodynamics, depend on the variables
$\theta$ and $\nc$ only through the ratio $\tN=\theta/\nc$. This is puzzling,
because the  effective action of QCD is periodic in $\theta$ with period
$2\pi$. The resolution of the paradox is given in 
ref.~\cite{Leutwyler Smilga}, where the relation between QCD
and gluodynamics is discussed in detail. The paradox is related to the fact
that the partition function 
$ Z_{\mbox{\tiny GD}}$ involves a sum over all gluon field configurations, 
while $Z_{\mbox{\tiny QCD}}$ only extends over those on which
fermions can live. In gluodynamics, the sum over all configurations
involves fractional winding numbers -- only the quantity $\nc\, \nu$ must be an
integer. In the presence of fermions, however, the boundary conditions imply 
that $\nu$ itself must be an integer. 
Even if their mass is sent to infinity, the fermions thus exert a
restriction on the gluon configurations to be summed over: Only configurations
with integer winding number contribute. This implies that, if either the
number of colours or the quark masses are sent to infinity, 
the QCD partition function does not in general tend to the partition function
of gluodynamics. Instead, it approaches the one obtained by
restricting the sum over all gluon configurations to
those with integer winding number, given by  
the projection  
\bea \hat{Z}_{\mbox{\tiny GD}}(\theta)=
\frac{1}{\nc}\sum_{k=0}^{\nc-1}Z_{\mbox{\tiny GD}}(\theta + 2 \pi k)\nonumber
\co 
\eea 
which removes fractional winding numbers. While $Z_{\mbox{\tiny GD}}(\theta)$ 
is periodic in $\theta$ with period $2 \pi \nc$, the projection 
$\hat{Z}_{\mbox{\tiny GD}}(\theta)$ has the
$2\pi$ periodicity characteristic of $Z_{\mbox{\tiny QCD}}(\theta)$.
If $\theta$ is taken constant and the fourdimensional volume $V$ is large, 
the partition function of gluodynamics is dominated by the contribution
from the ground state,
\bea Z_{\mbox{\tiny GD}}(\theta)=\exp \{-V \nc^2 e_0(\tN)\}\fs\eea
For $\nc\rightarrow \infty$, the partition function of QCD thus approaches
the expression
\bea Z_{\mbox{\tiny QCD}}(\theta)\rightarrow
\frac{1}{\nc}\sum_{k=0}^{\nc-1}\exp \{- V \nc^2 e_0({\tN}_k)\}\co\hspace{2em}
{\tN}_k=\frac{\theta + 2 \pi k}{\nc}\fs
\eea
The large V limit picks out the term in the sum for which
$e_0({\tN}_k)$ is minimal. This means that, in the large $\nc$
  limit, the vacuum energy density of QCD is determined by 
\bea \bar{e}_0=\mbox{min}\hspace{-0.8em}\rule[-0.7em]{0em}{0em}_k  
\hspace{0.7em}e_0({\tN}_k) \fs \nonumber \eea
Since a change in $\theta$ by $2\pi$ is equivalent to a shift in $k$, the
quantity $\bar{e}_0$ is periodic with period $2\pi$, while 
$e_0$ is periodic only with respect to $\theta\rightarrow \theta +2\pi \nc$.

The distinction between the large $\nc$ limit of QCD and gluodynamics only
matters for values of $\theta$ outside the interval $-\pi<\theta<\pi$. 
Within this region, we have $\bar{e}_0=e_0$. For all other values,
$\bar{e}_0$ represents the periodic continuation from that interval.
In our context, only the infinitesimal neighbourhood of the point 
$\theta=0$ matters -- we are using the external field $\theta(x)$ merely as
a technical device to analyze the properties of the Green functions for 
$\theta=0$. In that context, we do not need to distinguish 
between $e_0$ and $\bar{e}_0$.

We add a remark concerning the topological susceptibility. As discussed
in ref.~\cite{Minkowski}, the correlation function $\lvac T \omega(x)\, 
\omega(0)\rvac$ is too singular for the integral 
\bea \chi(p^2)=-i\!\!\int\!\! d^4\!x\, e^{i p \cdot x}\,\lvac T \omega(x)\, 
\omega(0)\rvac\nonumber \eea
to exist, in QCD as well as in gluodynamics.
The corresponding
dispersive representation contains two subtractions,
\bea \chi(p^2)= \chi(0)+p^2\chi'(0)+\frac{p^4}{\pi}\int_0^\infty\!\!
\frac{ds}{s^2(s-p^2)}\,\mbox{Im}\,\chi(s)\fs \nonumber\eea
Accordingly, if the susceptibility is defined as an integral
 over the correlation function, it is an ambiguous notion. We instead
identify it with the mean square winding number per unit volume,
$\chi(0)=\langle \nu^2\rangle /V$, which does not 
suffer from such an ambiguity. In particular, if the large $\nc$ limit is 
taken at nonzero quark masses, we have\footnote{The equality only holds if
  $\nc \qm$ is large compared to 
the scale of the theory. In the chiral limit, 
$\tau_{\rm\sss QCD}$ vanishes, so that $\tau_{\rm\sss QCD}\longrightarrow
\hspace{-1.4em}/ \hspace{1.4em}
\tau_{\rm\sss GD}$.} 
\bea \tau_{\rm\sss QCD} = \tau_{\rm\sss GD}+O(\nc^{-1})\co\hspace{2em}
\tau_{\rm\sss GD} = O (1)\fs\eea 
The ambiguity in the first derivative 
remains -- it reflects the fact that, in the presence of a
space-time-dependent vacuum angle, the Lagrangian of 
QCD must be supplemented with a contact term $\propto
D_\mu \theta D^\mu\theta$. A renormalization of the corresponding coupling 
constant is needed to absorb the quadratic divergence in the
correlation function $\lvac T \omega(x)\, \omega(0)\rvac$. An analogous term
also occurs at the level of the effective theory. 

\setcounter{equation}{0}
\section{Massless quarks}
In the present section, we briefly review some of the consequences of the U(1)
anomaly, because these play a central role in the analysis of the low energy
structure at large $\nc$. For simplicity, we consider massless quarks, but
leave the number of flavours open. 
We normalize the singlet axial current with
\bea A^0_\mu=\bar{q}\,\mbox{$\frac{1}{2}$}\lambda_0\gamma_\mu\gamma_5 \,q=
\frac{1}{\sqrt{2\nf}}\,\bar{q}\,\gamma_\mu\gamma_5 \,q\co\eea
where $\nf$ is the number of flavours. The
divergence of this current is given by 
\bea \label{anomalous WI}
\partial^\mu \!A_\mu^0=\sqrt{2 \nf}\, \omega\fs\eea
We write the $\eta'$ matrix elements in the form
\bea \lvac A_\mu^0\,|\eta'\rangle=
i\,p_\mu F_0\co\hspace{2em}
\lvac\omega|\eta'\rangle=\sqrt{2\nf}\,\frac{\tau}{F_0}\fs
\nonumber \eea
The relation (\ref{anomalous WI}) then shows that the mass of the $\eta'$ is
given by
\bea M_{\eta'}^2=\frac{2\nf\, \tau}{F_0^2}\co\eea
and the Ward identities obeyed by the correlation functions of $A_\mu^0$
and $\omega$ lead to the representation
\bea i\!\!\int\!\!dx\,e^{i p \cdot x}\,\lvac T A_\mu^0(x)\,
A_\nu^0(0)
\rvac\al=\al\frac{p_\mu p_\nu\,F_0^2 }{M_{\eta'}^2-p^2}+
g_{\mu\nu} F_0^2+g_{\mu\nu} R_0(p^2)\no\al\al 
\hspace{5em}+(p_\mu p_\nu-g_{\mu\nu}p^2)R_1(p^2)\co\no
\int\!\!dx\,e^{i p \cdot x}\,\lvac T A_\mu^0(x)\,\omega(0)
\rvac\al=\al\frac{p_\mu\,\sqrt{2\nf}\,\tau}{M_{\eta'}^2-p^2}+
\frac{1}{\sqrt{2\nf}}\,p_\mu R_0(p^2)\co \label{correlation functions}\\
\frac{1}{i}\!\int\!\!dx\,e^{i p \cdot x}\,\lvac T \omega(x)\,\omega(0)
\rvac\al =\al-\frac{2\nf\,\tau^2}{F_0^2\,(M_{\eta'}^2-p^2)}+
\tau-\frac{1}{2\nf}\,p^2 R_0(p^2)\fs\nonumber\eea
The large $\nc$ counting rules of section \ref{gluodynamics}  
imply that the correlation function of the axial current is a quantity
of $O(\nc)$. For the pole term contained therein to be consistent with this,
the decay constant $F_0$ must be of $O(\sqrt{\nc})$. Similarly, for the
pole term in the correlation function formed with $A^0_\mu(x)$ and $\omega(x)$
not to generate a contribution that diverges for $\nc\rightarrow\infty$, 
the constant $\tau$ must represent a quantity of $O(1)$.
In the large $\nc$ limit, the pole term in $\lvac T\omega(x)\,\omega(0)\rvac$
therefore disappears, as it should: That function must approach
the correlation function of gluodynamics. In this theory, the mass gap
persists when $\nc\rightarrow \infty$, so that the expansion in powers of 
the momentum is an ordinary Taylor series, which starts with the
topological susceptibility,
\bea \tau=\tau_{\mbox{\tiny GD}}+O(1/\nc)\co\hspace{2em}R_0(p^2)=O(1)\fs\eea
In QCD, the low energy structure is more intricate, particularly when
$\nc$ becomes large. In addition to the $\nf^2-1$ Goldstone bosons 
generated by the
spontaneous breakdown of SU($\nf$)$_\indR\times$SU($\nf$)$_\indL$, the spectrum contains
a further state that also becomes massless in the limit
$\nc\rightarrow\infty$. At large $\nc$, the low energy structure of the 
theory contains a new scale, $M_{\eta'}$, which is independent of the quark 
masses and of the intrinsic scale of QCD.
The magnitude of the pole terms in the above 
correlation functions, for instance, is sensitive to the relative size 
of the momentum $p$ compared to this scale.

\setcounter{equation}{0}
\section{Simultaneous expansion in \boldm{$p\!\!$} and \boldm{$1/\nc$}}
\label{simultaneous expansion}
In order to analyze 
the behaviour in the region where the momenta
are of the size of $M_{\eta'}$ -- in particular, on the mass shell of the
$\eta'$ -- we need to treat both $p$ and $M_{\eta'}$ as small quantities.
This can be done in a controlled manner in the framework of a 
simultaneous expansion in
powers of momenta and $1/\nc$, ordering the series
with \cite{bounds}
\bea\label{delta0} p=O(\sqrt{\delta})\co\hspace{2em}
1/ \nc=O(\delta)\fs\eea
In that bookkeeping, the denominator associated with $\eta'$-exchange
represents a small quantity of $O(\delta)$, so that the corresponding low
energy singularities are enhanced. In the preceding section, we considered
the $1/\nc$ expansion at fixed $p$ -- as noted there, the pole term in the
correlation function of $\omega(x)$ then represents a subleading contribution.
In the simultaneous expansion we are considering now, however, this term 
is of the same algebraic order as the
one from the topological susceptibility of gluodynamics. 
Moreover, the functions $R_0(p^2)$ and $R_1(p^2)$ are now expanded in powers 
of $p$. Since the leading terms are of order
$R_0(0)=O(1)$,
$R_1(0)=O(\nc)=O(1/\delta)$, the contributions from these functions are 
suppressed by one power of $\delta$ as compared to the pole terms:
For massless quarks,  
the leading term in the $\delta$-expansion of the three correlation functions
considered in the preceding section is obtained by simply dropping
the contributions from $R_0(p^2)$ and $R_1(p^2)$ and is thus 
fully determined by the low energy
constants $\tau_{\mbox{\tiny GD}}$ and $F_0$.

The result may be converted into a simple statement about the effective 
action. The correlation functions under discussion specify the part
of $S_{\eff}$ that is bilinear in the 
external fields $a_\mu^0(x)$ and $\theta(x)$. At leading order in the
$\delta$-expansion, this part only involves the constants
$\tau_{\mbox{\tiny GD}}$ and $F_0$.
The explicit expression reads
\bea S_{\eff}\al=\al\mbox{$\frac{1}{2}$}\!\! \int\!\! dx \,dy
\,f(x)f(y)\Delta_{\eta'}(x-y) + \mbox{$\frac{1}{2}$}\!\int\!\! dx
\{F_0^2a_\mu^0(x)\,a^{0\,\mu}(x)-\tau_{\mbox{\tiny GD}} \,\theta(x)^2\}+\ldots
\nonumber\eea
where $f(x)$ is a linear combination of the external fields 
$\partial^\mu a_\mu^0(x)$ and $\theta(x)$ and $\Delta_{\eta'}(x)$ is the
propagator of the $\eta'$: 
\begin{eqnarray*}
\hspace*{0.7em}f(x)= F_0\partial^\mu a_\mu^0(x)-\sqrt{2\nf}\,
\frac{\tau_{\mbox{\tiny GD}}}{F_0}\,
\theta(x)\co\hspace{0.9em}\Delta_{\eta'}(x)=\frac{1}{(2\pi)^4}\!
\int\!dp\,\frac{e^{-i p\cdot x}}{M_{\eta'}^2-p^2-i\epsilon}\fs
\end{eqnarray*}
The expression represents the classical action of a free scalar
field $\psi$ in the presence of external fields. The relevant Lagrangian is
given by
\bea\label{LU1} {\cal L}_{\psi}\al=\al\frac{F_0^2}{4\nf}
D_\mu\psi D^\mu\psi-
\frac{\tau_{\mbox{\tiny GD}}}{2}(\psi+\theta)^2\co\hspace{2em}
D_\mu\psi=\partial_\mu\psi -2\langle a_\mu\rangle\co\eea
where
$\langle a_\mu \rangle=a_\mu^0 \sqrt{\frac{1}{2}\nf}$ is the trace of the
axial external field. 
If we assign the fields the weight
\bea\theta(x)=O(1)\co\hspace{2em} 
a_\mu(x)=O(\sqrt{\delta})\co\hspace{2em}\psi(x)=O(1)\co\eea
all of the contributions occurring in this Lagrangian represent terms of 
$O(1)$. 

Above, we determined the effective Lagrangian from the properties of the
correlation functions. We could instead have investigated the general
expression permitted by the symmetries of the theory -- indeed, we will
make use of that method to work out the higher order terms in 
the $\delta$-expansion. At leading order, however, that approach
leads to the following problem, which arises from the presence of a new low
energy scale and does not occur in the effective theory relevant for finite
$\nc$. Let us switch the external fields off. Up to total derivatives,
the general Lorentz invariant expression quadratic in $\psi$ is of the form
\bea  {\cal L}=\sum_{n=0}^\infty c_n\,\psi\,\wave^n\psi+\ldots\nonumber
\eea
A priori, it is not legitimate to dismiss terms with more than two
derivatives: If the coefficient $c_n$ is of order
$\nc^n$, then it does contribute at the leading order of the 
$\delta$-expansion. There is a good reason, however, why such terms do not 
occur. The point is that the propagator of the field $\psi$ is given by the
inverse of the function $\sum_n c_n(-p^2)^n$. Unless this expression is
a second order polynomial in $p$, the propagator will thus have more than
one zero. With $c_n\propto \nc^n$, all of these occur in the region
where $p^2$ is small, of order $1/\nc$ -- the spectrum of the theory would
thus contain more than one particle that (i) has the quantum numbers of the 
$\eta'$ and (ii) becomes massless in the large $\nc$ limit. Our explicit
calculation of the pole terms in the correlation functions relies on the 
very plausible assumption that only one such particle occurs -- this is why
higher derivative terms do not occur at leading order in
$\delta$. 

\setcounter{equation}{0}
\section{Effective theory in the U(1) sector}\label{effective theory}
The result obtained in the preceding section shows that -- as far as the
two-point-functions of the operators $A_\mu^0(x)$ and $\omega(x)$
are concerned, and for massless quarks -- the leading terms in the 
simultaneous expansion in powers of momenta and $1/\nc$ can be characterized
by means of a remarkably simple effective field theory. It involves a single
dynamical variable $\psi(x)$, describing the degrees of freedom of the
$\eta'$. Before generalizing this result to the other correlation functions
and to nonzero quark masses, we add a few remarks about the structure 
of the effective Lagrangian obtained above.

First, we note that the Lagrangian (\ref{LU1}) is manifestly invariant 
under local U($\nf$)$_{\indR}\times$U($\nf$)$_{\indL}$ 
rotations,
provided the dynamical
variable is transformed according to
\bea\label{trafopsi} \psi'=\psi  -i\ln\det V_{\indR}+i\ln\det V_{\indL}\fs\eea
This ensures that the sum $\psi+\theta$ is invariant, so that the
same is true of the term proportional to $\tau_{\mbox{\tiny GD}}$. 
Since the trace of $a_\mu$
transforms like an Abelian gauge field, the term $D_\mu\psi$ also represents
an invariant. 

Next, we observe that the choice of the dynamical variable is not unique.
The normalization of the field $\psi$, for instance, can be chosen such
that the kinetic term takes the standard form
$\frac{1}{2}\,\partial_\mu\psi\,\partial^\mu\psi$ (the reason for not doing
so is that this would ruin the simplicity of the transformation law 
(\ref{trafopsi})).
More generally, the variable $\psi$ can be replaced by a function thereof,
which moreover may involve the derivatives of $\psi$ and the external fields
-- 
the result for the action remains the same. 

This implies that the effective
Lagrangian is not unique. An infinitesimal change of variables generates a
term in the Lagrangian that is proportional to the equation of
motion. Conversely, contributions proportional to the equation of motion
may always be removed from the Lagrangian by performing a suitable
transformation of variables. This freedom is characteristic of effective
theories, where the dynamical fields represent mere variables of
integration in the path integral and are not of physical interest. 
The transformation $\psi\rightarrow \psi +\kappa\cdot (\psi+\theta)$, for 
instance,
preserves the transformation law (\ref{trafopsi}), but takes the Lagrangian 
into
\bea\label{LU2} {{\cal L}_{\psi}}^{\prime}\al=\al
\frac{F_0^2}{4\nf}\left\{(1+\kappa)^2 D\psi^2
+2\kappa(1+\kappa)D\psi D\theta+\kappa^2 D\theta^2\right\}-
\frac{\tau_{\mbox{\tiny GD}}}{2}(1+\kappa)^2(\psi+\theta)^2,\nonumber\eea
with $D_\mu\theta=\partial_\mu\theta +2\langle a_\mu\rangle$. 
For this choice of variables, the Lagrangian contains additional terms, 
proportional to $D_\mu \psi D^\mu\theta$ and $D_\mu \theta D^\mu\theta$,
respectively. Note also that
the coefficient of the one $\propto (\psi +\theta)^2$ is then not given by 
the topological susceptibility of gluodynamics when $\nc$ becomes large.
In the following we stick to the form of the effective
Lagrangian in eq.~(\ref{LU1}).

The low energy analysis relies on the perturbative expansion of
the effective theory. At leading order, the
effective action is given by the classical action, which collects the
contributions from the tree graphs. The quantum fluctuations of the
effective field are treated as corrections. In the standard framework,
where the $\eta'$ does not occur among the effective degrees of freedom,
the size of the quantum fluctuations is controlled by the external momenta
and by the quark masses $\qm=O(p^2)$:
Graphs involving $\ell$ loops are suppressed by a factor of order
$p^{2\ell}$. In the extended framework, however, these fluctuations
involve three scales, the external momenta, the quark masses and the mass of 
the $\eta'$.
Their magnitude can be controlled algebraically only if $M_{\eta'}$
is also treated as small. For the 
consistency of the effective
framework, it is essential that the number of colours provides us with
an algebraic parameter that controls the size of
$M_{\eta'}=O(1/\sqrt{\nc})$. In fact, we will see that in the simultaneous 
expansion in powers 
of momenta, quark masses and $1/\nc$, the quantum fluctuations only
start contributing at order $\delta^2$.

The problems arising if $M_{\eta'}$ is not treated as a small parameter
can be seen already at tree level, where the effective action is obtained
by evaluating the classical action at its extremum. There, the field
$\psi(x)$ obeys the equation of motion, which is
of the form $\wave \psi + M_{\eta'}^2\psi=f$.
At leading order in an expansion in powers of $p$ at fixed
$M_{\eta'}^2$, this equation reduces to $\psi=f/M_{\eta'}^2+\ldots$
The kinetic term then only occurs as a correction. In the loop integrals,
the $p$-expansion replaces the propagators by the series
$(M_{\eta'}^2-p^2)^{-1}=1/M_{\eta'}^2+\ldots$ It is clear that those
integrals cannot be analyzed in this manner.

\setcounter{equation}{0}
\section{Extension to U(3)}
We now discuss the extension required to analyze the full
effective
action and set $\nf=3$. 
The dynamical variables of the effective theory must then account for
all particles that become massless in the limit $\nc\rightarrow \infty$,
$\qm\rightarrow 0$. In that limit, the
symmetry group of the Hamiltonian is
$\mbox{G}=\mbox{U(3)$_{\indR}\times$U(3)$_{\indL}$}$. 
We assume that this symmetry is spontaneously broken and that 
the ground state yields a nonzero
expectation value for the quark condensate $\lvac\qbar_{\ind L}q_{\indR}\!\rvac$.
While the subgroup generated by the vector charges remains intact 
\cite{Vafa Witten}, the U(1)$_{\ind A}$ rotations
generated by the singlet axial charge are spontaneously broken: The
operator $\qbar_{\ind L}q_{\indR}$ transforms in a nontrivial manner under
these. Accordingly, the ground state is 
symmetric only under the subgroup $\mbox{H}=\mbox{U(3)}_{\indV}$. The 
dynamical variables of the effective theory live on the coset space 
$\mbox{G}/\mbox{H}=\mbox{U(3)}$, so that we may collect the effective fields
in a matrix $U(x)\in\mbox{U(3)}$. The nine parameters of the coset space
correspond to the nine massless pseudoscalar fields needed to describe
the Goldstone bosons. 
The extension from the standard framework, where the effective field
is an element of SU(3), to the one we are considering here 
shows up in the phase of the determinant
\bea\label{detU} \det U(x)=e^{i\psi(x)}\fs\eea
The field $\psi(x)$ describes the $\eta'$. Under the action of the symmetry
group, $U(x)$ transforms with 
\bea \label{trafoU}U'(x)=V_{\indR}(x)U(x) V_{\indL}^\dagger(x)\fs\eea
The transformation law
(\ref{trafopsi}) represents a special case 
of this formula, obtained by comparing the determinants of the left and
right hand sides. In canonical coordinates, the matrix $U(x)$
is parametrized in terms of nine pseudoscalar fields $\phi^0(x)$, $\ldots$ ,
$\phi^8(x)$:  
\bea\label{Upi} U(x)=\exp\, i\, \sum_{k=0}^8\lambda_k\phi^k(x)\co
\eea
where $\lambda_1,\ldots ,\,\lambda_8$ are the Gell-Mann matrices and 
$\lambda_0=\sqrt{2/3}$. In these coordinates, the singlet field $\psi(x)$
is represented by  $\psi(x)=\sqrt{6}\,\phi^0(x)$. 

The calculation described in section \ref{simultaneous expansion} only
concerns the part of the effective Lagrangian 
that governs the dynamics of the $\eta'$ field $\psi(x)$. 
The part relevant for the
dynamics of the pseudoscalar octet may, however, be worked out in the same
manner. If the quark masses are set equal to zero,
the eightfold way is an exact symmetry, so that there is no mixing between 
octet and singlet. The leading term in the $\delta$-expansion of the
correlation function of the octet components of the axial current, 
$\lvac TA^i_\mu(x)\,A^k_\nu(0)\rvac$, is obtained along the same lines as in
the case of the singlet component.
The result may again be characterized
by an effective Lagrangian that is
quadratic in the corresponding dynamical variables -- the
fields $\phi^k(x)$ in this case:
\bea\label{Lpi} {\cal L}_{\phi}\al=\al\mbox{$\frac{1}{2}$}\,F^2\sum_{k=1}^8
(\partial_\mu\phi^k-a^k_\mu)(\partial^\mu\phi^k-a^{\mu\,k})\fs\eea
In the chiral limit, which we are considering, a mass term does not occur
here: The pseudoscalar octet is then strictly massless. The relevant part of
the effective Lagrangian contains a single effective coupling constant,
related to the matrix element
\bea \lvac A_\mu^i|\pi^k\rangle=i\, \delta^{ik} p_\mu F \fs\nonumber \eea
In fact, in the large $\nc$ limit, the correlation functions of the
singlet and octet currents become identical: The ratio $F_0/F$ tends to 1. 
Accordingly, the sum
${\cal L}={\cal  L}_\psi+{\cal L}_\phi$ takes the simple form
\bea\label{Lpsipi} {\cal L}\al=\al\mbox{$\frac{1}{2}$}\,F^2\sum_{k=0}^8
(\partial_\mu\phi^k-a^k_\mu)(\partial^\mu\phi^k-a^{\mu\,k})
-\frac{\tau_{\mbox{\tiny GD}}}{2}(\psi+\theta)^2\fs\eea

\setcounter{equation}{0}
\section{Chiral symmetry}\label{chiral symmetry}
There is an essential difference between the Lagrangians relevant
for the octet and singlet degrees of freedom: While ${\cal L}_{\psi}$ 
is invariant under chiral
transformations, ${\cal L}_{\phi}$ is not. The difference arises 
from the fact that the part of the chiral group that matters for 
${\cal L}_{\psi}$ is the Abelian U(1) factor, while the one that
counts for ${\cal L}_{\phi}$ is a nonabelian group, 
SU(3)$_{\indR}\times$SU(3)$_{\indL}$. This implies that
the pseudoscalar octet mesons necessarily interact among themselves: The Ward
identities for the various Green functions intertwine these with one another, 
so that a nonvanishing two-point-function can occur only together with
Green functions containing more than two currents. In contrast to the
one in eq.~(\ref{LU1}), the Lagrangians in eqs.~(\ref{Lpi}) and (\ref{Lpsipi}) 
cannot stand by themselves.

The interaction terms required by chiral symmetry are well known. 
They are generated automatically, if the term
$\partial_\mu\phi^k-a_\mu^k$ is replaced by the covariant derivative of the 
effective field. For the matrix $U(x)$ that includes the $\eta'$, the
covariant derivative is given by
\bea D_\mu U\al=\al \partial_\mu U-i\,(v_\mu+a_\mu)\hspace{0.06em}U +i\,
U(v_\mu-a_\mu)\fs\eea 
Under chiral rotations, this object transforms in the same manner as the
field $U(x)$ itself. The expansion in
powers of the meson fields thus starts with 
\bdm D_\mu U=i\,\sum_{k=0}^8(\partial_\mu \phi^k - a_\mu^k)\,\lambda^k
+\ldots\edm 
Hence, the Lagrangian (\ref{Lpsipi}) represents the quadratic term in
the chirally invariant expression (as usual, $\langle A\rangle$ stands for 
the trace of $A$)
\bea\label{L00} {\cal L}=\mbox{$\frac{1}{4}$}F^2\langle D_\mu U^\dagger D^\mu
U\rangle -\frac{\tau_{\mbox{\tiny GD}}}{2}(\psi+\theta)^2
\fs\eea
The above calculation demonstrates that this Lagrangian 
correctly characterizes the two-point-functions formed with the winding 
number density and the singlet and
octet components of the axial current -- at leading order of the
$\delta$-expansion and in the chiral limit. In fact, in this limit,
the Lagrangian (\ref{L00}) properly accounts for the leading contributions 
in the $\delta$-expansion of all of the Green functions that can be formed
with these operators. For a proof, we refer to appendix A.

We may decompose the effective field into a part that only contains
the $\eta'$ and a part that describes
the pseudoscalar octet:
\bea\label{dec} U= e^{\frac{i}{3}\psi} \hat{U}\co\hspace{2em}
\mbox{det}\,\hat{U}=1\fs\eea
It is convenient to define the covariant derivative of $\hat{U}\in
\mbox{SU(3)}$ by
\bea D_\mu \hat{U}=\partial_\mu \hat{U}-i\,(\hat{v}_\mu
+\hat{a}_\mu)\hspace{0.06em}\hat{U} +i\,\hat{U}(\hat{v}_\mu
-\hat{a}_\mu)\co\eea
where $\hat{v}_\mu$, $\hat{a}_\mu$ are the traceless parts of
$v_\mu$, $a_\mu$. By construction, this quantity obeys $\langle \hat{U}^\dagger
D_\mu\hat{U}\rangle =0$, to be compared with
$\langle U^\dagger D_\mu U\rangle= iD_\mu \psi$ for  
$U\in\mbox{U(3)}$. The relation between the two derivatives,
\bea D_\mu U=e^{\frac{i}{3}\psi}\left\{D_\mu\hat{U}+
\mbox{$\frac{i}{3}$}\,D_\mu\psi\,\hat{U}\right\}\co\nonumber\eea
implies the identity
\bea   \langle D_\mu U^\dagger D^\mu U \rangle=
\langle D_\mu \hat{U}^\dagger D^\mu \hat{U} \rangle+
\mbox{$\frac{1}{3}$}
D_\mu\psi D^\mu\psi\co\nonumber\eea
which splits the above Lagrangian into
an SU(3) part that exclusively involves the Goldstone boson octet and 
a U(1) part that only contains the singlet meson field $\psi(x)$. 
The singlet axial field and the vacuum angle only occur in the U(1) part, 
while the vector and axial octets
only appear in the SU(3) part -- the  singlet vector field does not enter at
all. For massless quarks, the octet and singlet
sectors thus separate like oil and water. 

\setcounter{equation}{0}
\section{Full effective Lagrangian}\label{full Lagrangian}
For nonzero quark masses, the properties of the theory may be analyzed in 
terms of an expansion in powers of these, so that we are then dealing
with a triple expansion in (i) the number of derivatives, (ii) powers of quark 
masses and (iii) powers of $1/\nc$. It is convenient to generalize the
ordering (\ref{delta0}) used for the massless theory by counting the quark
masses as quantities of order $\delta$:
\bea\label{delta1} \partial_\mu
=O(\sqrt{\delta}\;)\co\hspace{2em}m=O(\delta)\co\hspace{2em}1/\nc=O(\delta)\fs
 \eea
In this bookkeeping, the vacuum angle 
and the effective fields $U(x)$, $\psi(x)$ are treated as 
quantities of order 1, while the external fields $v_\mu(x)$, $a_\mu(x)$,
$s(x)$, $p(x)$  
count as small perturbations of the same order as the derivatives
and the quark masses, respectively: 
\bea\label{delta2} (U,\,\psi,\,\theta)=O(1)\co\hspace{2em}(v_\mu,\,a_\mu)=
O(\sqrt{\delta}\;)\co\hspace{2em}(s,\,p)=O(\delta)\fs\eea

Collecting the contributions of order $1,\,\delta,\,\delta^2,\ldots$, the 
full effective Lagrangian takes the form (see appendix \ref{appendix A}): 
\bea {\cal L}_{\eff}=  
{\cal  L}^{(0)}+{\cal L}^{(1)} +{\cal L}^{(2)}+  
 \ldots\eea
The individual terms only contain a finite number of effective coupling
constants. In particular, the leading term ${\cal L}^{(0)}=O(\delta^0)$ 
exclusively involves the coupling constants $F,B$ and $\tau$. Replacing the
external fields $s(x)$ and $p(x)$ by
\bea\label{chi} 
\chi(x)\equiv 2B\left\{s(x)+ip(x)\right\}\co\eea the explicit expression reads
\bea\label{L0}
{\cal L}^{(0)}=\mbox{$\frac{1}{4}$}F^2\langle D_\mu U^\dagger D^\mu U\rangle +
\mbox{$\frac{1}{4}$} F^2\langle U^\dagger \chi + \chi^\dagger U \rangle -
\mbox{$\frac{1}{2}$}\,\tau (\psi + \theta)^2\fs\eea

Note that ${\cal L}^{(0)}$
contains vertices of the type $F^2\partial
\phi\partial\phi\phi^{n-2}$, which represent interactions among the
pseudoscalar mesons. The corresponding tree graph contribution to the
scattering amplitude describes a
collision with altogether $n$ particles in the initial and final states.
In the large $\nc$ limit, the contribution disappears, in proportion to
$F^{2-n}\propto\nc^{1-n/2}$, in accordance with the general counting of 
powers for the scattering amplitude given in appendix \ref{appendix A}. When
$\nc$ is sent to infinity, all of the bound states become free particles of 
zero width. The larger the number of particles participating in the 
reaction, the smaller the interaction among these.
 
In the large $\nc$ limit, diagrams with a
single quark loop dominate, so that only contributions with a single 
trace over flavour matrices occur (OZI rule). This property entails that the
two-point-function of the singlet current approaches the one
of the octet components, in particular $F_0/F\rightarrow 1$. In general,
however, the matrix elements of the singlet and octet components are
different, even if $\nc$ is taken large. In particular, the OZI rule does
not imply that the four-point-function of the singlet axial
current
approaches the one of the octet components -- in fact, it does not:
The latter contains a four-fold pole, whose residue is proportional
to the scattering amplitude. The term represents a contribution of 
order $\nc$, which at low energies is fully determined by the pion decay 
constant. An analogous contribution
to the four-point-function of the singlet axial current does not occur,
because the scattering amplitude 
$\eta'\eta'\rightarrow\eta'\eta'$ does not pick up a term proportional to
$1/F^2$: Crossing symmetry implies that, for massless quarks, 
the coefficient is proportional to  $s+t+u=4M_{\eta'}^2=O(1/\nc)$.

The term ${\cal L}^{(1)}=O(\delta)$ contains the contributions of 
$O(\nc\, p^4)$, $O(p^2)$ and $O(1/\nc)$.  
The first may be copied from the 
SU(3) Lagrangian, simply dropping those coupling constants
that violate the OZI-rule. Concerning the remaining contributions, the 
terms that are of the same structure as those in 
${\cal L}^{(0)}$ may be absorbed in the coupling constants $F,B,\tau$ (note
that the coefficient of the term proportional to $(\psi+\theta)^2$ then
differs from the topological susceptibility of gluodynamics -- we need to
distinguish the coupling constant $\tau$ from $\tau_{\mbox{\tiny GD}}$).
With a suitable choice of the dynamical variables, the explicit expression 
for ${\cal L}^{(1)}$ can be brought to the form
\bea\label{L1} {\cal L}^{(1)}\al=\al L_2\langle D_\mu U^\dagger D_\nu U 
D^\mu U^\dagger D^\nu U\rangle +(2L_2+L_3) \langle D_\mu U^\dagger D^\mu U 
D_\nu U^\dagger D^\nu U\rangle\no\al\al+
L_5 \langle D_\mu U^\dagger D^\mu U (
U^\dagger \chi  +\chi^\dagger U )\rangle +
L_8\langle U^\dagger \chi U^\dagger\chi +\chi^\dagger U \chi^\dagger U\rangle
\no\al\al-i L_9\langle R_{\mu\nu} D^\mu U D^\nu U^\dagger+
L_{\mu\nu} D^\mu U^\dagger D^\nu U\rangle+L_{10} \langle
R_{\mu\nu}UL^{\mu\nu} U^\dagger\rangle
\\\al\al+\mbox{$\frac{1}{12}$}F^2\Lambda_1D_\mu \psi D^\mu\psi
-\mbox{$\frac{1}{12}$}F^2 \Lambda_2 \,i(\psi+\theta)
\langle U^\dagger\chi- \chi^\dagger U\rangle \no\al\al +
\mbox{$\frac{1}{12}$}H_0 D_\mu\theta D^\mu\theta 
+H_1\langle R_{\mu\nu} R^{\mu\nu}+
L_{\mu\nu} L^{\mu\nu}\rangle+H_2\langle\chi^\dagger \chi\rangle
\fs\nonumber\eea
The covariant derivatives and the field strength tensors are defined by
\bea
D_\mu U\al=\al \partial_\mu U-i\,(v_\mu+a_\mu)\hspace{0.06em} U +i\,
U (v_\mu-a_\mu) \co \non \\ 
D_\mu\psi\al=\al\partial_\mu\psi -2\langle a_\mu\rangle\co\hspace{4.9em}
D_\mu\theta =\partial_\mu\theta +2\langle a_\mu\rangle\co  \\ 
R_{\mu \nu } \al=\al \partial_\mu r_\nu - \partial_\nu r_\mu - i
[r_\mu,r_\nu]\;\co\;\;  
L_{\mu \nu } = \partial_\mu l_\nu - \partial_\nu l_\mu - i
[l_\mu,l_\nu]\co
\non 
\label{notation}\eea
where $r_\mu = v_\mu + a_\mu$ and $l_\mu = v_\mu - a_\mu$. 
The somewhat queer numbering of the coupling constants arises because we
retain the notation introduced in ref.~\cite{GL SU(3)} to denote the 
couplings at first nonleading order.  The
SU(3) Lagrangian given in that reference contains three
independent terms with four derivatives, involving the coupling constants
$L_1$, $L_2$ and $L_3$, respectively. For $\nc\rightarrow \infty$,
all of these are of order $\nc$, but the OZI rule suppresses the
combination $2L_1-L_2$. The above expression for those terms that involve
four derivatives is obtained from the Lagrangian of ref.~\cite{GL SU(3)}
by replacing the constant $L_1$ by $\frac{1}{2}L_2$. 
The counting rules imply that the coupling constants $L_2,L_3,L_5,L_8,L_9,
L_{10}$ represent quantities of $O(\nc)$, while 
$\Lambda_1$, $\Lambda_2$ are of $O(1/\nc)$. Concerning the contact terms,
$H_0$ is of order 1, while $H_1$ and $H_2$ are of $O(\nc)$. 

The number of independent coupling constants entering the low energy 
representation
of the effective action to first nonleading order 
is roughly the same as for the SU(3) Lagrangian:
While the four parameters $L_1-\frac{1}{2}L_2,L_4,L_6,L_7$ are relegated to 
next-to-next-to leading order, three new coupling constants appear: 
$\tau,\Lambda_1,\Lambda_2$. The framework allows us to also evaluate 
the correlation functions
formed with the operators $A^0_\mu$ and $\omega$, while the Lagrangian
of ref.~\cite{GL SU(3)} accounts for these
only at leading order.
A further virtue of a Lagrangian that
explicitly includes the degrees 
of freedom of the $\eta'$ is that it equips the contributions from
$\eta'$-exchange with the proper denominator and, for instance, distinguishes
the factor $1/M_{\eta'}^2$ from $1/(M_{\eta'}^2-M_\eta^2)$. In view of
$M_\eta^2/M_{\eta'}^2=0.33$, this distinction is numerically quite
significant. On the other hand, the above framework retains only the 
next-to-leading terms of the $1/\nc$ expansion. 
In sections \ref{higher} and \ref{comparison}, we will consider those
contributions of 
next-to-next-to leading order that are needed to fully match
the SU(3) and U(3) theories on their common domain of validity.

\setcounter{equation}{0}
\section{Potentials}
At this point, we establish contact with the notation introduced in
ref.~\cite{GL SU(3)}, where the vertices of the
effective Lagrangian are ordered in the same 
fashion as for the case of SU(3), that is by simply counting powers
of momenta and quark masses. We refer to that ordering as the $p$-expansion.
Ordering the various couplings in this manner,
the Lagrangian takes the form
\bea\label{LV} {\cal L}_{\eff}\al=\al  -V_0+
V_1\,\langle D_\mu U^\dagger D^\mu U\rangle
+V_2\,\langle U^\dagger \chi \rangle+V_2^\star\,
 \langle\chi^\dagger U \rangle\\\al\al
+V_3\,D_\mu\psi D^\mu\psi +
 V_4\, D_\mu \psi D^\mu\theta +
V_5 \,D_\mu\theta D^\mu\theta+O(p^4)\fs\nonumber\eea
The leading term $V_0$ contains those vertices that do not involve 
derivatives or quark mass factors. Chiral symmetry implies that the 
matrix $U$ cannot appear without derivatives, but it does not protect the 
combination $\psi+\theta$ of the singlet field and the vacuum angle. 
There are vertices without derivatives containing  
an arbitrary power of this combination. Their collection $V_0$ 
represents a function of the variable $\psi+\theta$. The same applies for the 
coefficients of those contributions that do involve quark mass factors
or derivatives. The effective
coupling constants are the coefficients $V_{n,\,k}$ occurring in their
expansion in powers of $\psi+\theta$,
\bea V_n=\sum_{k=0}^\infty V_{n,\,k}\,(\psi+\theta)^k\fs\nonumber\eea
The functions $V_n(\psi+\theta)$  may be viewed as potentials that describe the
dynamics in the U(1) sector, with $V_0$ as the most important one. 
The counting rules established in appendix
\ref{appendix A} imply that -- for a suitable choice of the dynamical
variables -- we have
\bea \al\al V_{0,\,k}= O(\nc^{2-k})\co\hspace{1em}
\{ V_{1,\,k},\,V_{2,\,k}\}= O(\nc^{1-k})\co\hspace{1em}
 \{V_{3,\,k},V_{4,\,k},V_{5,\,k}\}=O(\nc^{-k})\fs\nonumber\eea
Indeed, these rules were written down already in ref.~\cite{GL SU(3)}. 
The detailed analysis described in the present paper merely fills a gap: 
The argumentation given in that reference is incomplete, because it does 
not take the ambiguities into account that arise from the freedom in the
choice of the dynamical variables. As demonstrated in the appendix, 
these ambiguities are in one-to-one correspondence with those occurring in the
off-shell extrapolation of the matrix elements. What we have shown now is that 
there exists a class of such extrapolations, for which the above
counting rules for the effective coupling constants strictly follow from those
for the correlation functions. We will make use of this result 
when analyzing the Kaplan-Manohar transformation.

At leading order, the freedom in the choice of the dynamical variables
is hidden in the manner in which the field $U(x)$ is parametrized in 
terms of the variables $\phi^0(x),\ldots,\phi^8(x)$ -- that is in the
choice of the coordinates on the group U(3). As we are expressing the
Lagrangian in terms of $U(x)$, it is irrelevant how these coordinates 
are chosen. At first nonleading order, however, the ambiguities inherent in 
the choice of the dynamical variables do show up and the Lagrangian takes
a unique form only if that freedom is fixed. As discussed in appendix
\ref{appendix A}, this can
be done by eliminating invariants that vanish if the field $U(x)$ obeys the
equation of motion associated with ${\cal L}^{(0)}$. The representation
for ${\cal L}^{(1)}$ in eq.~(\ref{L1}) fully exploits this freedom. The 
eleven couplings occurring there are both complete and independent: 
With a suitable change of variables, all of the invariants that can be built 
up to and including $O(\delta)$ reduce to a linear combination of those 
listed and, conversely,
the choice made fixes the dynamical variables -- up to transformations of
$O(\delta^2)$. In the notation introduced above, the choice of variables made
in eq.~(\ref{L1}) corresponds to $V_4=0$,
as far as the terms of $O(\delta)$ are concerned. One readily checks that
a suitable change of variables of the type $U\rightarrow e^{if}U$, where
$f$ only depends on $\psi+\theta$, removes $V_4$ to all orders and thereby
fixes the dynamical variables up to transformations 
of $O(p^2)$.

At order $p^4$, the natural parity part of the effective Lagrangian
contains altogether 52 potentials. For an explicit representation,
we refer to \cite{Herrera 1}. The potentials relevant for the 
unnatural parity part are given explicitly in section \ref{unnatural ho}. 
In connection with the Kaplan-Manohar
transformation, the symmetry breaking terms of order $m^2$ are of
special interest. These are
of the same structure as in the case of the SU(3) Lagrangian:
\bea\label{Lm} {\cal L}_{\chi^2}\al=\al 
W_1\langle U^\dagger\chi U^\dagger \chi\rangle+ 
W_1^\star\langle \chi^\dagger U \chi^\dagger U\rangle+
W_2\langle U^\dagger \chi \rangle^2\label{Lagm}\\ \al\al
+W_2^\star\langle \chi^\dagger U\rangle^2+
W_3\langle  U^\dagger \chi\rangle\langle \chi^\dagger U\rangle+W_4\langle
\chi^\dagger \chi \rangle\fs\nonumber\eea
The only difference is that the coupling constants $L_6,L_7,L_8$ and $H_2$  
are replaced by the potentials $W_1,\ldots,W_4$, which depend on the
variable $\psi+\theta$.

In the $1/\nc$ expansion, most of the potentials occurring in ${\cal L}_2$
start contributing only at next-to-next-to leading order: If only
terms of order $\delta$ are retained, the effective Lagrangian 
reduces to the explicit expression in eq.~(\ref{L1}). 

\setcounter{equation}{0}
\section{Dependence on the vacuum angle \boldm{$\theta$}}\label{dep theta} 
 
Since the dependence of the various quantities on the vacuum angle is rather
peculiar, particularly in the large $\nc$ limit, we now briefly discuss
this issue. Throughout this section we restrict ourselves to a constant value
of the external field $\theta(x)=\theta$, turn the pseudoscalar 
field $p(x)$ off and identify the scalar one with the quark mass matrix,
$s(x)=m$.

The transformation law (\ref{trafo}) shows that a suitable global U(1) 
transformation, for instance 
$V_{\indR}=e^{\frac{i}{3}\theta}{\bf 1}$, $V_{\indL}={\bf 1}$, removes the
vacuum angle. The external vector and axial fields stay put, but the
quark mass matrix undergoes a change: $m\rightarrow m_\theta$. 
The invariance of the effective action thus
implies 
\bea\label{Sinv} 
S\{v,\,a,\,\theta,\,m\}=S\{v,\,a,\,0,\,m_\theta\}\co\hspace{2em}
m_\theta=e^{\frac{i}{3}\theta} m
\co\eea
the familiar statement that the vacuum angle only enters in combination with
the quark mass matrix: Only the phase $\mbox{arg\,det}\, m_\theta=
\mbox{arg\,det}\,  m +\theta$ matters.

In the effective Lagrangian, however, the vacuum angle does not enter
in this combination. Also, according to eq.~(\ref{trafoU}), the effective
field undergoes a change under the above
transformation: $U\rightarrow e^{\frac{i}{3}\theta}U$. We need to be careful 
when performing finite transformations on the effective field. In the
construction of the effective Lagrangian, only the series in powers of
both the external fields and the dynamical variables $\phi(x)$ were considered.
If we now take a vacuum angle of finite size, we are summing the expansion
in powers of $\theta$ to all orders. This may lead to ambiguities. The
relation $\mbox{det}\,U=e^{i\psi}$, for instance, can be solved for $\psi$ only
up to a multiple of $2\pi$. This means that the decomposition in
eq.~(\ref{dec}) is unique only in the vicinity of $U=1$. 

Since the effective theory is evaluated perturbatively, its properties are
governed by the tree graphs, that is by those of the corresponding 
classical field theory. In this framework, ambiguities of the type
just mentioned are avoided if $\psi$ is replaced by the gauge invariant 
field 
$\tpsi=\psi+\theta$, collecting 
the remaining dynamical variables in the matrix
$\Ubar$:
\bea\label{Ubar} U=e^{\frac{i}{3}\tpsi}\,\Ubar\co\hspace{2em}
\tpsi=\psi+\theta\co\hspace{2em}\mbox{det}\,
\Ubar=e^{-i\theta}\fs\eea
It is convenient to define the covariant derivative of $\Ubar$ by 
\bea \label{DUbar}\al\al D_\mu\Ubar=\partial_\mu \Ubar -i\,
(v_\mu+\bar{a}_\mu)
\Ubar+i\,\Ubar (v_\mu-\bar{a}_\mu)\co\\
\al\al\bar{a}_\mu=a_\mu-
\mbox{$\frac{1}{3}$}\langle a_\mu\rangle -\mbox{$\frac{1}{6}$}\partial_\mu
\theta=a_\mu-\mbox{$\frac{1}{6}$}D_\mu
\theta\co\nonumber\eea
so that $\langle \Ubar^\dagger D_\mu \Ubar\rangle=0$. In this notation,
the derivative of $U$ reads
\bea
D_\mu U=e^{\frac{i}{3}\tpsi}\left\{D_\mu
  \Ubar+\mbox{$\frac{i}{3}$}(\partial_\mu\tpsi-D_\mu\theta)\Ubar
\right\}\co \eea
and the leading term in the effective Lagrangian takes the form
\bea 
{\cal L}^{(0)}\al=\al\mbox{$\frac{1}{4}$}F^2\langle D_\mu \Ubar^\dagger 
D^\mu \Ubar + e^{-\frac{i}{3}\tpsi}\Ubar^\dagger
\chi  +
 e^{\frac{i}{3}\tpsi} \chi^\dagger\Ubar \rangle
+\mbox{$\frac{1}{12}$} F^2 (\partial_\mu\tpsi-D_\mu\theta)^2-
\mbox{$\frac{1}{2}$}\tau \tpsi^2
\fs\nonumber\eea
The equation of motion belonging to this
Lagrangian unambiguously determines the classical
solution $\tpsi_{cl}$, $\Ubar_{cl}$ in terms of the external fields.
If these are subject to a gauge transformation, 
$\tpsi_{cl}$ remains invariant, while $\Ubar_{cl}$ goes into
$V_{\indR}\Ubar_{cl}V_{\indL}^\dagger$, even if the angles of the
transformation are taken large. This also holds
if the higher order contributions to the effective Lagrangian are accounted
for. In particular, under the global U(1) rotation
considered above, which maps the vacuum angle into zero,
the classical solution does transform with
$\Ubar_{cl}\rightarrow e^{\frac{i}{3}\theta}\Ubar_{cl}$. 

At leading order
of the perturbative expansion, the effective action is given by the
value of the classical action at the extremum.
Hence it does obey the relation (\ref{Sinv}), also if $\theta$ is taken 
large. Note that, once the dynamical variables of the effective theory are
identified with $\tpsi$ and $\Ubar$, the effective Lagrangian 
contains the external field $\theta(x)$
exclusively through the derivatives thereof -- a constant vacuum angle 
then exclusively manifests itself
via the constraint $\mbox{det}\,\Ubar=e^{-i\theta}$. This also implies
that the effective theory automatically generates an effective action 
that is periodic in the vacuum angle: The only place where a constant shift
in that angle shows up is through the factor $e^{-i\,\theta}$, which 
remains the same if $\theta$ is replaced by $\theta+2\pi$.

The dependence on $\theta$ disappears in the large $\nc$ limit, both in 
gluodynamics and in QCD. The vacuum energy density, for instance, 
represents a term of $O(\nc^2)$, but
a dependence on $\theta$ only shows up at $O(1)$, through the
susceptibility term. The origin of this suppression 
can immediately be seen in the structure of the QCD Lagrangian: 
The properties of the theory are governed by the term 
$g^{-2}G_{\mu\nu}G^{\mu\nu}$, while the vacuum angle enters through
$\theta \,G_{\mu\nu}\tilde{G}^{\mu\nu}$. In the large $\nc$ limit, the 
term containing the vacuum angle is smaller than the one that
determines the dynamics by the factor $g^2\theta=O(1/\nc)$.  

Gauge invariance implies that, in the effective Lagrangian, 
the vacuum angle only appears in the combination $\tpsi=\psi+\theta$,
together with the singlet field. In particular, the term proportional 
to $\theta^2$ in the vacuum energy density is converted into one 
proportional to $\tpsi^2$ and thus equips the $\eta'$ with a mass.
That mass disappears in the large $\nc$ limit, because the 
$\theta$-dependence of the vacuum energy of gluodynamics is suppressed.

Small quark masses also suppress the dependence on $\theta$.
If the determinant of the quark mass matrix vanishes, the Green functions
of the vector and axial currents and hence also the scattering matrix 
elements even become entirely independent of $\theta$. Concerning the
$\theta$-dependence, the quark masses -- the parameters that break chiral 
symmetry -- thus play a similar role
as the parameter $1/\nc$ that measures the breaking of the OZI rule.

As an immediate corollary of the fact that, in the large $\nc$ limit,
the dependence on the vacuum angle is suppressed, the potentials
reduce to constants in that limit. More generally, if the expansion in 
$1/\nc$ is cut off at a finite order, the potentials are replaced 
by polynomials in $\tpsi$. 

\setcounter{equation}{0}
\section{Kaplan-Manohar transformation at large \boldm{$\nc$}}
\label{KaplanManohar}
As pointed out by Kaplan and Manohar \cite{Kaplan Manohar}, the
standard SU(3) Lagrangian is invariant under the transformation in
eq.~(\ref{KMintro}), 
provided the effective coupling constants $L_6,L_7$ and $L_8$ are subject to a
corresponding change. In that framework, the vacuum angle is set to zero.
The transformation can, however, be generalized to allow for nonzero values
of $\theta(x)$. To ensure that the modified mass matrix
has the same transformation properties as the original one also with respect
to chiral U(1)-rotations, it suffices to equip the transformation law
for the scalar and pseudoscalar external fields with a factor
of $e^{-i\theta}$. With $m(x)=s(x)+ip(x)$, the Kaplan-Manohar transformation
then takes 
the form
\bea\label{KM} m\rightarrow m + \lambda \,e^{-i\theta}\,
m^{\dagger\,-1}\det m^\dagger\fs\eea

In the present context, it is convenient to order the vertices with the
$p$-expansion, using the representation of the effective
Lagrangian in eq.~(\ref{LV}). At order $p^2$, the scalar and pseudoscalar
external fields only enter trough the 
term $V_2\langle  U^\dagger \chi\rangle$ and its complex conjugate, with
$\chi=2 B m$. Under the above operation, this term picks up a
contribution proportional to 
$\langle U^\dagger \chi^{\dagger\,-1}\rangle\det \chi^\dagger$.
The expression may be simplified by applying the identity
\bea  \langle C^{-1}\rangle= \frac{\langle C\rangle^2-
\langle C^2\rangle}{2\,\det C}\nonumber\eea
to the matrix $C=\chi^\dagger U$. In view of $\det U=e^{i\psi}$, the
transformation then yields
\bea V_2\langle U^\dagger \chi  \rangle\rightarrow
V_2\langle U^\dagger \chi 
\rangle+\frac{\lambda}{4B}\,V_2\,e^{-i(\psi+\theta)}
\left\{\langle \chi^\dagger U\rangle^2
-\langle \chi^\dagger U  \chi^\dagger  U \rangle\right\}\fs\nonumber\eea
The modification may be absorbed in a suitable change of the
potentials that describe the
mass terms contained in the effective Lagrangian of order $p^4$, which
are given in eq.~(\ref{Lm}).
Indeed, the effective Lagrangian does remain invariant under the transformation
(\ref{KM}), provided the effective coupling constants contained in
the potentials $W_1$ and $W_2$ are modified according 
to 
\bea \al\al W_1\rightarrow W_1+\frac{\lambda}{4B}\,
V^\star_2\,e^{i(\psi+\theta)}\label{W1W2}\co\\
\al\al W_2\rightarrow W_2-\frac{\lambda}{4B}\,
V^\star_2\,e^{i(\psi+\theta)}\co\nonumber\eea
all other coupling constants staying put. 
This demonstrates that the general U(3) Lagrangian exhibits the same kind of
reparametrization invariance as the one built on SU(3).

The modification of the potentials $W_1$, $W_2$, however, is in conflict
with the behaviour of the theory at large $\nc$. As discussed in detail in the
preceding section, the effective Lagrangian represents a polynomial in 
the variable $\psi +\theta$, at any order of the
$1/\nc$ expansion -- an immediate consequence of the fact that
the dependence of the matrix elements on the vacuum angle is suppressed
in the large $\nc$ limit. In view of the factor $e^{i(\psi+\theta)}$,
the changes required in $W_1$ and $W_2$ violate this condition, 
already at leading order. The disaster evidently originates
in the phase factor $e^{-i\theta}$ occurring in the Kaplan-Manohar
transformation (\ref{KM}):
This factor is needed for the modified quark mass matrix
to transform in the same manner as the physical one, but it introduces
a dependence on $\theta$ that is in conflict with the large $\nc$ properties
of the theory.

It does not come as a surprise that the Kaplan-Manohar transformation breaks
 the 
 Okubo-Iizuka-Zweig rule: The transformation mixes different quark flavours.
It is to be expected that the parameter $\lambda$ can at most be 
of order $1/\nc$. This also follows from the fact that the 
coupling constant $L_6$ picks up a term proportional to $\lambda F^2$.
Since the counting rules imply that $L_6$ represents a term of $O(1)$,
while $F^2$ is of order $\nc$, the transformation can be consistent with the
large $\nc$ properties of the theory only if $\lambda$ represents a term
of order $1/\nc$. The above analysis reaches much further: It shows that
the parameter $\lambda$ vanishes to all orders of the $1/\nc$ expansion.

\setcounter{equation}{0}
\section{Renormalization of the operators \boldm{$A_\mu^0\!\!$} and
 \boldm{$\omega$}} 
\label{anomalous}
As is well-known, the dimension of the singlet axial current is anomalous
\cite{Kodaira}.  The operator
$A^0_\mu=\qbar\frac{1}{2}\lambda_0\gamma_\mu\gamma_5q$ must be renormalized
for the correlation functions formed with this current to remain finite
when
the cutoff is removed.  In particular, the one particle matrix elements $\lvac
A_\mu^0|P \rangle=ip_\mu F_P^0$ require renormalization: The singlet decay
constants $F_{\pi^0}^0$, $F_\eta^0$, $F_{\eta'}^0$ represent quantities like
the quark masses or the quark condensate -- they must be renormalized. 

In the
$\overline{\mbox{MS}}$ scheme, the renormalized quantities depend on a running
scale. To distinguish this scale from the one used to renormalize the loop
graphs of the effective theory, we denote it by $\mu_{\QCD}$.  There is a
difference in that the scale dependence of the quark masses shows up already
at leading order in the $1/\nc$ expansion, while the anomalous dimension of
the singlet axial current only manifests itself at next-to-leading order,
because the triangle graph responsible for the phenomenon contains an extra
quark loop.

The representation of the 
renormalization group only interrelates operators with the same 
dimension and the same Lorentz quantum numbers. 
Moreover, the matrix elements of the 
representation exclusively involve those coupling constants that are 
dimensionless in the dimension of physical interest -- 
the QCD coupling constant $g$ and 
the vacuum angle $\theta$ in the present case. In particular, the
representation is independent of the quark masses. If these are turned off,
the charges of the flavour group SU(3)$_{\indR}\times$SU(3)$_{\indL}$ are 
conserved.
The corresponding representation on the set of field operators commutes
with the one of the renormalization group. 

Since the axial singlet current does not have any partners with
the same Lorentz and flavour quantum numbers and the same dimension,
it transforms as a singlet under the renormalization group. 
The operators $\qbar_{\indR}^{\,i} q_{\indL}^k$
form an irreducible representation of the flavour group. As the theory does
not contain any other operators with these quantum numbers and with the same
dimension, the renormalization is multiplicative also in this case,
\bea A_\mu^{0\,ren}=\za A_\mu^0\co\hspace{2em} 
(\qbar_{\indR}^{\,i} q_{\indL}^k)^{ren}=\zm\, \qbar_{\indR}^{\,i}
q_{\indL}^k\fs\eea
The renormalization of the singlet decay constants, for instance, reads
\bea F_P^{0\,ren}=\za
F_P^0\co\hspace{2em}P=\pi^0,\,\eta\,,\,\eta'\fs 
\nonumber\eea
The renormalization factors depend
on the running scale of QCD:
   \begin{eqnarray}
   \mu_{\QCD} \frac{d \za }{d \mu_{\QCD}} = \ga \za \co \spc
   \ga  = - \frac{6 \nf (\nc^2-1)}{\nc}\left(\frac{g}{4
   \pi}\right)^4 + 
   O(g^6) \fs \label{anodim1}  
   \end{eqnarray}
\begin{eqnarray}
   \mu_{\QCD} \frac{d \zm }{d \mu_{\QCD}} = \gm \zm \co \spc
   \gm  = \frac{3(\nc^2-1)}{\nc}\left(\frac{g}{4
   \pi}\right)^2  + 
   O(g^4) \fs \label{anodim2}  
   \end{eqnarray}

The renormalization of the operator $\omega$ is more complicated. As
this field represents the variable conjugate to $\theta$, the issue is related
to the dependence of the effective action on the vacuum angle.
This dependence is of crucial importance for the analysis of the theory
at large $\nc$. For this reason, we now discuss it in detail. 
  
The external fields may be viewed as space-time dependent coupling constants.
The renormalization of the operators amounts to a renormalization of these
``constants''. In particular, if the vacuum angle is turned off, 
the effective action remains the same
if we replace the bare operators  $\qbar_{\indR}^{\,i} q_{\indL}^k$,
$\qbar_{\indL}^{\,i} q_{\indR}^k$ and $A_\mu^0$ by
the renormalized ones and at the same time replace the bare external 
fields by the quantities $s^{\,ren}= \zm^{-1}s$, $p^{\,ren}= \zm^{-1} p$,
$\langle a_{\mu}\rangle^{ren}= \za^{-1}\langle a_\mu\rangle$, while the vector
field and the octet components of the axial field remain put -- the
corresponding operators do not get renormalized.
 
To see what happens if the vacuum angle $\theta$ is turned on, we exploit
the fact that the natural parity part of the effective action is invariant
under the local
U(3)$_{\indR}\times$U(3)$_{\indL}$ transformations specified in 
eq.~(\ref{trafo}).
We denote the octet part
of the axial field by $\hat{a}_\mu\equiv 
a_\mu-\frac{1}{3}\langle a_\mu\rangle$ and replace $s$, $p$ by
the combination
\bea   m_{\indtheta} = e^{\frac{i}{3}\theta}(s+ip) \co\eea 
which is invariant under the
transformations generated by the singlet axial charge.
Invariance under this subgroup then implies
that $\langle a_\mu\rangle$ and $\theta$ can enter the effective action
only through the gauge
invariant combination
$D_\mu \theta=\partial_\mu \theta + 2\langle a_\mu\rangle$,
\bea S_{\eff}=S_{\eff}\{v_\mu,\hat{a}_\mu,D_\mu\theta,m_{\indtheta}\}\eea
This shows that the functional $S_\eff$ is known for 
arbitrary $\theta(x)$ if it
is known for $\theta(x)=0$: The general expression is obtained from the one
relevant for $\theta(x)=0$ by replacing $\langle a_\mu\rangle $,
$s$ and $p$  with 
$\frac{1}{2}D_\mu\theta$, $\frac{1}{2}(m_{\indtheta}+m_{\indtheta}^\dagger)$
and $\frac{i}{2}(m_{\indtheta}^\dagger-m_{\indtheta})$, respectively.
In particular, the quantities $D_\mu\theta$ and $m_\theta$ are renormalized
according to
\bea\label{ren} 
(D_\mu\theta)^{ren}=\za^{-1} D_\mu\theta\co\hspace{2em}
m_{\indtheta}^{\,ren}=\zm^{-1} m_{\indtheta}\fs \eea

\setcounter{equation}{0}
\section{Renormalization of the effective action}
\label{renormalization of the effective action}
For the effective action to remain finite when the cutoff is removed,
the Lagrangian must include all
terms of mass dimension less than or equal to four that are consistent
with the symmetries of the theory. This also applies to contact terms,
such as $h_0D_\mu\theta D^\mu\theta$, 
$h_1\langle R_{\mu\nu}R^{\mu\nu}+ 
L_{\mu\nu} L^{\mu\nu}\rangle$ or  $h_2\langle m^\dagger m
\rangle$. The constant $h_1$, for instance, 
is needed to renormalize the QCD contributions to the
electric charge. It generates contact terms in the correlation functions
of the vector and axial currents.   
Together with the renormalization of the bare coupling constant $g$, 
the operation ensures that all of the renormalized correlation functions 
formed with the
quark currents and with $\omega$ approach finite limits,
\bea\label{rg}\hspace*{-1em} 
S_\eff\{v_\mu,\hat{a}_\mu,D_\mu\theta,m_{\indtheta},g,h,\mu_0\}\!=\!
S_\eff\{v_\mu,\hat{a}_\mu,(D_\mu\theta)^{ren},m_{\indtheta}^{ren},g^{ren},
h^{ren},\mu_{\QCD}\}.\;
\eea
We have denoted the coupling constants associated with the
contact terms collectively by $h$ and use the $\overline{\mbox{MS}}$ scheme
for the renormalized quantities. These are independent
of the cutoff $\mu_0$, but do depend on the running scale $\mu_{\QCD}$. 
In the effective action, the running scale drops
out -- the left hand side of the above relation is scale independent.

A priori, since
$\theta$ is dimensionless, the representation of the renormalization group
might depend on this coupling constant, too. As discussed above, however,
$\theta$ is an inessential parameter in the sense that the manner in
which the effective action depends on it is fully determined by the
symmetries of the theory. In particular, the
factors $\za$ and $\zm$ are the same as for $\theta=0$. 
In fact, the renormalization law for $m_\theta$
shows that the vacuum angle is a scale independent quantity -- in agreement
with the fact that the effective action is periodic in this variable.

The renormalization law for $D_\mu \theta$, on the other hand,
implies that the renormalization of the singlet axial field is 
rather complicated: The effective action approaches a finite
limit only if, in addition to a multiplicative renormalization with the factor
$\za^{-1}$, the field
$\langle a_\mu\rangle$ is simultaneously also 
subject to a U(1) gauge transformation by the angle 
$\frac{1}{2}(\za^{-1}-1)\,\theta(x)$,
\bea \langle a_\mu\rangle^{ren}=\za^{-1}\langle
a_\mu\rangle+\mbox{$\frac{1}{2}$}\,(\za^{-1}-1)\,\partial_\mu\theta \fs\eea
The relation concisely specifies the
renormalization of the correlation functions involving the operator 
$\omega$. In contrast to  
$A_\mu^0$ and $\qbar_{\indR}^{\,i}
q_{\indL}^k$, for which the renormalization is multiplicative, $\omega$ 
is subject to an inhomogeneous renormalization, despite the fact that
the variable conjugate to it, the vacuum angle, is 
renormalization group invariant.

The origin of the complication is readily understood: It reflects the
conservation law of the singlet axial current 
\bea \partial^\mu \! A^0_\mu= \sigma^0
+\sqrt{6}\,\omega
\co\hspace{2em}\sigma^0=\sqrt{\mbox{$\frac{2}{3}$}}\;\qbar \,i\gamma_5 \qm  q
\eea
(for simplicity, 
we identify $s$ with the
quark mass matrix $\qm$ and set $p=0$).
While $\partial^\mu \! A^0_\mu$ picks up a factor of $\za$, the term
$\sigma^0$ is invariant, 
because the quark mass matrix transforms contragrediently to the operator 
$\qbar\, i\gamma_5 q$. The matrix
element $\lvac \omega|\eta'\rangle$, for instance, 
may be represented as a difference of
two contributions. The first is proportional to $M_{\eta'}^2F_{\eta'}^0$
and thus scales with $\za$, while the second is given by the matrix
element $\lvac\sigma^0\, |\eta'\rangle$, which does not get renormalized.

In contrast to the operator $\omega$ itself, the zero momentum projection
thereof, the winding number $\nu=\int\!dx\,\omega$, is renormalization group
invariant. The correlation functions of this quantity may be obtained from the
effective action by considering a constant vacuum angle and taking
derivatives with respect to it. The term proportional to $\partial_\mu\theta$
occurring in the renormalization of $\langle a_\mu\rangle$ is then absent.
Since the vacuum angle is invariant under renormalization, the same holds
for the correlation functions of $\nu$.

To illustrate the statement, consider the topological
susceptibility of QCD
\bea \tau_{\QCD}=-i\!\!\int\!\! dx\lvac T \omega(x)\, \omega(0)\rvac\fs
\nonumber \eea
If $s$ is identified with the quark mass matrix, the vacuum angle is taken
constant and all other external fields are switched 
off, the effective action 
reduces to $S_\eff=-\int\!dx\, \epsilon(\qm,\theta)$, where 
$\epsilon(\qm,\theta)$ is the vacuum energy density of QCD. 
The topological susceptibility is the second
derivative of $\epsilon(\qm,\theta)$ with respect to  
$\theta$. It can be represented in terms of the correlation function
$\lvac T \sigma^0(x)\,\sigma^0(y)\rvac$ and the quark condensate, by invoking 
the Ward identities
\bea \partial^\mu\lvac TA_\mu^0(x) \omega(y)\rvac\al=\al\lvac T\sigma^0(x)
\omega(y)\rvac
+ \sqrt{6}\,\lvac T\omega(x)\,\omega(y)\rvac\no
\partial^\mu\lvac TA_\mu^0(x)\, \sigma^0(y)\rvac\al=\al\lvac T\sigma^0(x)\,
\sigma^0(y)\rvac
+ \sqrt{6}\,\lvac T\omega(x)\,\sigma^0(y)\rvac\no
\al\al-i\,\mbox{$\frac{2}{3}$}\delta(x-y)\lvac\qbar\,\qm q\rvac\nonumber
\fs\eea
Since the left hand sides represent total derivatives, they drop out when 
taking the
integral over all of space. This leads to the representation 
\begin{eqnarray*} \hspace*{0.2em}\tau_{\QCD}=\frac{i}{\!\!\sqrt{6}}
\int\!\!dx \lvac T \omega(x)\,\sigma^0(0)\rvac=-\frac{i}{6}
\int\!\!dx\lvac T \sigma^0(x)\,\sigma^0(0)\rvac-\frac{1}{9}\,
\lvac \qbar\, \qm \,q\rvac.\nonumber
\end{eqnarray*}
The relation confirms the statement that the susceptibility is 
renormalization group invariant:
Neither the correlation function $\lvac T \sigma^0(x)\,\sigma^0(0)\rvac$ nor
the term involving the quark condensate, $\lvac \qbar\, \qm q\rvac$, 
depend on the running scale of QCD. 

\setcounter{equation}{0}
\section{Dependence of the effective coupling constants on the running scale
of QCD} \label{coupling}
We now translate these properties of the effective action of QCD 
into the language of the effective theory. The fact that the operators
$\qbar^{\,i}_{\indR}q^k_{\indL}$, $A_\mu^0$ and $\omega$ must be 
renormalized implies that some of the effective
coupling constants depend on the running scale of QCD. 
Apart from contact terms, the renormalization is fully determined by the
factors $\za$ and $\zm$ -- the coupling constant $g$ is hidden in the
effective couplings. The contact terms $h_0,h_1,\ldots$ of the
QCD Lagrangian are absorbed in the coupling constants $H_0,H_1,\ldots$
of the effective theory. The renormalization of these couplings 
thus involves the renormalization factors relevant for 
$h_0,h_1,\ldots$ In the following, we disregard the contact terms altogether
-- they lead a life of their own.
Note that the present section concerns 
the scale dependence of the effective coupling constants that arises 
from the renormalization of 
the QCD Lagrangian. The one generated by the logarithmic divergences
that occur within the effective theory is an entirely different issue 
(see section \ref{higher}). 

If the external fields 
$\langle a_\mu\rangle$ and $\theta$ are switched off, the renormalization
exclusively concerns $s$ and $p$. For the generating functional
of the effective theory relevant at fixed $\nc$ to become renormalization 
group invariant, the
low energy constant $B$ must be renormalized with 
$B^{ren}=\zm B$. The scale dependence of this constant 
cancels the one of the fields $s(x)$ and $p(x)$, so 
that the quantity $\chi=2B(s+ip)$ is independent of the QCD scale.
If the effective SU(3) Lagrangian is written in terms of this variable 
\cite{GL SU(3)}, all of the coupling constants occurring therein,
the pion decay constant in particular, are invariant under the
renormalization group.  

Let us now consider the extension to U(3) and include the external
fields $\langle a_\mu\rangle$ and $\theta$, so that the natural parity part of
the effective action becomes invariant under local 
U(3)$_{\indR}\times$U(3)$_{\indL}$. 
The renormalization properties discussed in the preceding section do not rely 
on the large $\nc$ limit. It is therefore appropriate to return to the 
form of the effective Lagrangian in eq.~(\ref{LV}), which does not
invoke the $1/\nc$ expansion. Inserting the decomposition (\ref{Ubar}),
we obtain ($V_4=0$):
\bea {\cal L}_{\eff}\al=\al-V_0+
V_1\langle D_\mu \Ubar^\dagger D^\mu \Ubar\rangle +V_2
\,e^{-\frac{i}{3}\tpsi}\langle
\Ubar^\dagger \chi \rangle+V_2^\star
\,e^{\frac{i}{3}\tpsi}\langle
\chi^\dagger \Ubar\rangle
\no\al\al+ (\mbox{$\frac{1}{3}$}V_1+V_3)( \partial_\mu\tpsi-D_\mu\theta)^2+ 
V_5 D_\mu\theta D^\mu\theta+O(p^4)\fs\eea
The covariant derivative $D_\mu\Ubar$ involves the octet components of 
the vector and axial fields, as well as the vacuum angle. Neither one of these
quantities undergoes renormalization, but the singlet
axial field in $D_\mu\theta =\partial_\mu\theta+2\langle a_\mu\rangle$ does:
In addition to a multiplicative renormalization, it picks up a U(1) gauge 
transformation proportional to $\partial_\mu\theta$, so that
$(D_\mu \theta)^{ren}=\za^{-1}D_\mu\theta$.
For the operation to map the effective Lagrangian onto itself, 
the variable $\tpsi$ must be renormalized in the same manner:
\bea \tpsi^{ren}=\za^{-1}\tpsi \co\label{renpsi}\eea
while $\Ubar$ remains put.
The Lagrangian then remains invariant, provided the effective
coupling constants are renormalized in such a manner that the potentials
transform with
\bea V_0(x)^{ren}\al=\al V_0(\za\,x)\no
V_1(x)^{ren}\al=\al V_1(\za\,x)\no
V_2(x)^{ren}\al=\al V_2(\za\,x)\,e^{-\frac{i}{3}(\za-1)x}\label{Vren}\\
V_3(x)^{ren}\al=\al \za^2\, V_3(\za x)+\mbox{$\frac{1}{3}$}(\za^2-1)
V_1(\za x)\no
V_5(x)^{ren}\al=\al \za^2\, V_5(\za x)\fs\nonumber\eea

We conclude that the effective 
Lagrangian is invariant under the renormalization of the external fields
specified in section \ref{renormalization of the effective action}, 
provided the dynamical variable $U$ is subject to renormalization: While the
unimodular part remains invariant, the
phase $\det U=e^{i\psi}$ transforms with
\bea \psi^{ren}=\za^{-1}\psi +(\za^{-1}-1)\theta\co\eea
Expressed in terms of $U$, the renormalization thus amounts to
\bea U^{ren}=e^{i\frac{z}{3}  \theta}\,
(\det U)^{\frac{z}{3}}\,U\co\hspace{2em}z
=\za^{-1}-1\fs\eea 

We repeat that we are disregarding contact terms. The above effective
Lagrangian does contain such a term:  $H_0=12 V_5(0)$. The renormalization of
$H_0$ involves the one of the QCD counter term $h_0$ and is not covered
by the above renormalization prescription for the potential $V_5$.  
Except for $H_0$, the relations (\ref{Vren}) do 
specify the renormalizations of all of the effective 
coupling constants occurring to order $p^2$, in terms of the factors
$\za$ and $\zm$ that characterize the anomalous dimensions of the operators
$A_\mu^0$ and $\qbar^{\,i}_{\indR}q^k_{\indL}$.
The couplings collected in the potential $V_0$, for instance,
are renormalized according to
\bea V_{0,\,k}^{ren}=\za^{k}V_{0,\,k}\fs\nonumber\eea
For the constant $\tau\equiv 2\,V_{02}$, this yields 
\bea \tau^{ren}=\za^2\tau\fs\eea 
Note that the factor $\za$ differs from 1 only by a term of order $1/\nc$.
As the triangle graph responsible for $\za$ does not occur in gluodynamics,
$\tau_{\mbox{\tiny GD}}$ does not get renormalized. 

The transformation property of $V_1=\frac{1}{4}F^2+O(x^2)$ shows that 
$F$ is independent of the QCD scale -- in agreement with the fact that
$F$ represents the value of a physical quantity in the chiral
limit. The constant $B$ transforms contragrediently to the quark mass
matrix, $B^{ren}=\zm B$, so that the lowest order mass formula 
$M_\pi^2=(m_u+m_d)B+\ldots$ does yield a scale independent pion mass.
For the coupling constants $\Lambda_1$ and $\Lambda_2$ the above
relations yield
\bea\label{renLambda} 
1+\Lambda_1^{ren}=\za^2(1+\Lambda_1)\co\hspace{1.5em}1+\Lambda_2^{ren}=
\za(1+\Lambda_2)\fs\eea
The result shows that the renormalization of the
effective coupling constants is in general not multiplicative.
This reflects the fact that the individual terms of the 
series ${\cal L}^{(0)}+{\cal L}^{(1)}+\ldots$ are not invariant under the
renormalization group. The leading term for instance contains the contribution
$\frac{1}{12}F^2 D_\mu\psi D^\mu\psi$,
which picks up a factor of $\za^{-2}$. A term proportional to 
$ D_\mu\psi D^\mu\psi$ also occurs
in ${\cal L}^{(1)}$. The two contributions add up to
$\frac{1}{12}F^2(1+\Lambda_1) D_\mu\psi D^\mu\psi$ -- the renormalization
of $\Lambda_1$ indeed ensures that the sum is
renormalization group invariant. Likewise,
the renormalization of the potential $ V_2$ generates a term 
proportional to $\za^{-1}-1$, which is absorbed in the
renormalization of $\Lambda_2$. 
The renormalization group thus intertwines
terms in the effective Lagrangian that involve the same number of derivatives
and quark mass factors, but carry a different power of $\nc$. 
This should barely come as a surprise -- the renormalization 
factor $\za$ itself represents an effect of the type $1+O(1/\nc)$. 
The renormalization of the nonleading couplings 
is not multiplicative, because it must cure the
deficiencies of the leading terms, so that the 
results obtained on the basis of  
the first two terms in the $1/\nc$ expansion of the 
effective Lagrangian do become renormalization group invariant
to first nonleading order.

The remaining coupling
constants of ${\cal L}^{(1)}$ are renormalization
group invariant, because the renormalization of ${\cal L}^{(0)}$ does not
generate terms of order $p^4$. They do, however, give rise to specific
nonmultiplicative renormalizations of the couplings occurring in ${\cal
  L}^{(2)}$ (see section \ref{higher}). 
 
Strictly speaking, 
the preceding discussion only shows that the
renormalization of the effective coupling constants which we have just given
is sufficient for the effective action to be 
independent of the QCD scale. One may also show that this renormalization
is necessary --  it suffices to 
calculate a few observables within the effective theory. The scale
independence of the result indeed implies that the effective coupling constants
must be renormalized in the above fashion \cite{Adelaide}.

\setcounter{equation}{0}
\section{Higher orders and loops}\label{higher}

There is a significant difference between the U(3) framework considered in
the present paper and the standard one, where the degrees of freedom of the
meson field are restricted to those of SU(3). In that case, 
the loop graphs of the effective theory are relevant
already at first nonleading order. Now, they only matter if we wish
to extend the calculation beyond this order.
As far as powers of momenta are concerned, the series is of
the same type in the two cases: Momenta count like 
$p\sim \sqrt{\delta}$. The loop graphs, however, are inversely proportional to 
powers of $F_\pi\sim\sqrt{\nc}$. While in the standard chiral 
perturbation series,
graphs containing $\ell$ loops generate contributions of order $p^{2\ell}$, 
they now only manifest themselves at order $\delta^{2\ell}$. In particular,
the one loop graphs yield contributions of next-to-next-to leading order.

Numerically, the one loop graphs are of about the same size as those 
occurring in the chiral perturbation series of SU(3) -- to count $F_\pi$
as a term of order $\sqrt{\nc}$ does not change the numerical value of
this coupling constant. The one loop graphs are relevant also if we wish to
establish the relation between the coupling constants occurring in 
the two versions
of the effective theory: The SU(3) coupling constants depend on the running 
scale used to renormalize these -- we
cannot make a significant comparison if we ignore the loops. 
The evaluation of the one loop graphs, however, poses a problem: 
Since these
violate the OZI rule, the divergences can be absorbed in a renormalization
of the effective coupling constants only if we include terms of $O(p^4)$ that 
are subleading in $1/\nc$. 

A complete analysis of the contributions of
next-to-next-to leading order 
is beyond our scope. The corresponding part of the effective Lagrangian
contains a plethora of terms of order $\nc\, p^6$, $p^4$, $p^2/\nc$ and
$1/\nc^2$, respectively. The first category contains vertices of the same 
structure as those occurring in the general SU(3) Lagrangian of order $p^6$, 
which are listed in ref.~\cite{SU3 order6}. For a complete set of terms
of order $p^4$, we refer to \cite{Herrera 1}.  
As the one loop graphs of ${\cal L}^{(0)}$
represent contributions that involve at most four powers of momentum, their
renormalization only requires counter terms that are at most of order
$p^4$. In other words, the coupling constants of the type $\nc\,p^6$ are
independent of the running scale of the effective theory. 
Also, since the QCD renormalization group preserves the Lorentz structure 
of the various vertices, it can intertwine the couplings of ${\cal L}^{(0)}$ 
and ${\cal L}^{(1)}$ only with those terms in ${\cal L}^{(2)}$ that are at 
most of order $p^4$.
In the following, we restrict ourselves to a small subset of the couplings
occurring at next-to-next-to leading order, the minimal set needed
to satisfy the following requirements:

(a) All of the couplings relevant 
for those quantities that can be calculated within the framework of
ref.~\cite{GL SU(3)} are included. This allows us to match the
two versions of the effective theory at one loop level and to express all of 
the SU(3) coupling constants in terms of those occurring in the
U(3) Lagrangian.

(b) All contributions needed to absorb the divergences occurring in the one 
loop graphs for the masses and decay constants of the
pseudoscalar nonet
are included, so that we can unambiguously account for
those contributions of next-to-next-to leading order
that are enhanced by a chiral logarithm. The terms required by (a)
suffice to satisfy this condition, except for one extra coupling. 
We retain the numbering
introduced in ref.~\cite{Herrera 1} and denote this term by $L_{18}$.

(c) The conditions (a) and (b) ensure that the result for the
{\em decay constants} is invariant under the renormalization group of QCD 
(see section \ref{coupling}). For this to be the case, it is essential
that the coupling constant 
$L_{18}$ is included: 
The renormalization group intertwines $L_5$ with this term. 
Likewise it intertwines $L_8$ with a further extra coupling
constant, $L_{25}$, which needs to be included 
for the {\em masses} of the pseudoscalars 
to also become renormalization invariant.

The effective Lagrangian then contains the following couplings:
\bea
{\cal L}^{(2)}\al=\al 
(L_1-\mbox{$\frac{1}{2}$}L_2)\langle D_\mu U^\dagger D^\mu U\rangle^2 +
L_4 \langle D_\mu U^\dagger D^\mu U\rangle
\langle U^\dagger \chi+ \chi^\dagger U \rangle\nonumber\\\al\al+
L_6\langle U^\dagger \chi+ \chi^\dagger U \rangle^2
+L_7\langle U^\dagger \chi - \chi^\dagger U \rangle^2
\label{Ldelta2}\\\al\al
+L_{18}\,iD_\mu \psi\,\langle D^\mu U^\dagger \chi-D^\mu U\chi^\dagger\rangle
+L_{25}\,i(\psi+\theta)\,\langle U^\dagger\chi U^\dagger\chi-\chi^\dagger
U\chi^\dagger U\rangle
\nonumber\fs\eea
As all of these terms represent contributions of $O(p^4)$, the coefficients
approach a finite limit when $\nc\rightarrow \infty$.
The contributions involving $L_1-\frac{1}{2}L_2,L_4,L_6$
and $L_7$ correspond to those pieces of the SU(3) Lagrangian that violate
the OZI-rule. 

The divergences generated by the one loop graphs of ${\cal L}^{(0)}$ 
have been worked out in
ref.~\cite{Herrera 1}. Expressed in terms of the factor
\bea
\lambda =\frac{\mu^{d-4}}{16\pi^2}\left\{\frac{1}{d-4}-\frac{1}{2}
\left(\rule[1em]{0em}{0em}\ln 4\pi+\Gamma'(1)+1\right)\right\}\eea
the renormalization of the effective coupling constants required to
absorb these divergences takes the form:
 \bea B\al=\al B^r\!\!\left\{1+\frac{4\tau}{F^4}\lambda \right\}\co\hspace{2em}
 L_n= L_n^r+\Gamma_n\lambda\co\hspace{2em}H_n=H_n^r+
\Delta_n\lambda\fs\eea
The constants $F$, $\tau$, $\Lambda_1 $ and $\Lambda_2$ do not get
renormalized.
The main difference to the case of SU(3) is that loops involving the
propagation of an $\eta'$ require a renormalization of the low energy 
constant $B$. The same graphs also generate a change in the values of the  
coefficients $\Gamma_6$, $\Gamma_8$ and $\Delta_2$:
\bea \Gamma_1\al=\al\mbox{$\frac{3}{32}$},\;\;\Gamma_2=\mbox{$\frac{3}{16}$},
\;\;\Gamma_3=0,\;\;
\Gamma_4=\mbox{$\frac{1}{8}$},\;\;\Gamma_5 =\mbox{$\frac{3}{8}$},\;\;
\Gamma_6=\mbox{$\frac{1}{16}$},\;\;\Gamma_7=0,\;\; \Gamma_8=
\mbox{$\frac{3}{16}$},\no\Gamma_9\al=\al\mbox{$\frac{1}{4}$},\;\;
\Gamma_{10}=-\mbox{$\frac{1}{4}$},\;\;
\Gamma_{18}= -\mbox{$\frac{1}{4}$},\;\;\Gamma_{25}=0,\;\;
\Delta_0=0,\;\;\Delta_1=-\mbox{$\frac{1}{8}$},\;\;\Delta_2=
\mbox{$\frac{3}{8}$}\fs\nonumber\eea

These relations concern the renormalization of the effective coupling 
constants required to absorb the divergences occurring within the effective 
theory -- the dependence on the renormalization scale of QCD is a 
different matter. For the coupling constants of ${\cal L}^{(0)}$
and ${\cal L}^{(1)}$, we have discussed the issue in detail
in section \ref{coupling}. That analysis is readily extended to the terms
contained in ${\cal L}^{(2)}$. We
only give the result: (a) The coupling constants $L_1,L_4,L_6$ and $L_7$
are independent of the QCD scale, also in the framework of U(3). 
(b) The constants 
$L_{18}$ and $L_{25}$ must be renormalized
according to
\bea 2L_5+3L_{18}^{ren}= \za (2L_5+3L_{18})\co\hspace{2em}
2L_8-3L_{25}^{ren}= \za (2L_8-3L_{25})\fs\eea
This demonstrates that both $L_{18}$ and $L_{25}$ are needed 
to arrive at a renormalization group invariant 
formulation of the effective theory at
next-to-next-to leading order. 

\setcounter{equation}{0}
\section{Matching U(3) and SU(3)}
\label{comparison}
The SU(3) framework applies if the quark masses and the external momenta 
are taken small compared to the mass generated by the topological 
susceptibility. Quantitatively, the condition takes the form \cite{Montpellier}
\bea m_s\,|\lvac\hspace{0.05em}\ubar\hspace{0.08em} 
u\rvac|\ll 9\, \tau_{\mbox{\tiny GD}}\fs \label{mlltau}\eea 
In this region, we may use the straightforward expansion in powers
of momenta and quark masses also for the U(3) theory, so that the two 
effective descriptions have a common region of validity, 
on which we can compare them.

As explicitly demonstrated in ref.~\cite{Herrera 4}, the path integral
for the effective U(3) Lagrangian indeed reduces to the one of
SU(3) if the singlet meson field is integrated out. The calculation
is analogous to the one described in \cite{GL SU(3)}, where the SU(3)
effective theory is matched
with the one relevant for SU(2). Like in that case, loops involving
the propagation of light as well as heavy mesons require special attention,
because the momentum scale of these is set by the heavy masses, while in the
light sector, that scale does not occur. We briefly sketch the essential
steps, referring to \cite{Herrera 4} for a more detailed discussion. 

To match the two effective theories, we again use the decomposition 
introduced in section \ref{dep theta}:
$U=e^{i\frac{1}{3}\tpsi}\,\bar{U},\;\tpsi=\psi+\theta$.
The phase factor explicitly exhibits the dependence of $U$ on the singlet 
field  $\psi$ and converts the relation $\det U=e^{i\psi}$ into the constraint
$\det \bar{U}=e^{-i\theta}$. The mapping ensures that, under chiral
rotations, $\bar{U}$ transforms in the same manner as $U$.
Expressed in terms of these variables, the first term in ${\cal L}^{(0)}$
takes the form
\bea \langle D_\mu U^\dagger D^\mu U\rangle =\langle D_\mu \Ubar^\dagger D^\mu
\Ubar\rangle+\mbox{$\frac{1}{3}$} (\partial_\mu \tpsi-D_\mu\theta)
(\partial^\mu \tpsi- D^\mu\theta)\fs
\nonumber\eea 
In the common domain of validity of the expansions in powers of $p$ and 
$\delta$, which is characterized by the inequality (\ref{mlltau}), 
the equation of motion implies that the field $\tpsi$ represents a
term of order $p^2$, so that we may expand the expressions in powers
of $\tpsi$. 

At leading order,
the matching reduces to a comparison of the classical actions.
Collecting
the relevant pieces of ${\cal L}^{(0)}$ and ${\cal L}^{(1)}$, we
find that the effective U(3) Lagrangian reduces to
\bea
{\cal L}_\eff\al =\al\mbox{$\frac{1}{4}$}F^2\langle D_\mu \bar{U}^\dagger
D^\mu  
\bar{U}\rangle + 
\mbox{$\frac{1}{4}$} F^2\langle   
\bar{U}^\dagger \chi+ \chi^\dagger
\bar{U} \rangle  \no
\al\al +\mbox{$\frac{1}{12}$}\left\{H_0 + F^2(1+ \Lambda_1)\right\}
D_\mu \theta D^\mu \theta 
+ O(p^4)\fs \nonumber\eea  
In the notation used here, the leading order SU(3) Lagrangian of ref.~\cite{GL
  SU(3)} reads\footnote{In ref.~\cite{GL SU(3)}, the covariant derivative
is defined as $\nabla_\mu \Ubar=\partial_\mu \Ubar +i\,(v_\mu+a_\mu)\Ubar
-i\,\Ubar(v_\mu-a_\mu)$. In view of eq.~(\ref{DUbar}), this amounts to 
  $\nabla_\mu\Ubar=D_\mu\Ubar-\frac{i}{3}D_\mu\theta \,\Ubar$.}
\bea\label{LSU3} \hspace*{-2em}{\cal L}^{\SU}_{p^2}
\al =\al\mbox{$\frac{1}{4}$}F^2\langle D_\mu \bar{U}^\dagger D^\mu 
\bar{U}\rangle + 
\mbox{$\frac{1}{4}$} F^2\langle   
\bar{U}^\dagger \chi+ \chi^\dagger
\bar{U} \rangle  
+\mbox{$\frac{1}{12}$} (\HSU_0+F^2)
D_\mu \theta D^\mu \theta \,.\eea  
Hence the two theories match at leading order, provided 
the coupling constants $F$ and $B$ are the same in both versions and the
  couplings $H_0$ are related by
\bea
\HSU_0 \al = \al  H_0 + F^2 \Lambda_1 +O(1/\nc)\fs  \eea

In order to match the coupling constants of next-to-leading order, we need to
perform the integration over the field $\tpsi$, which describes the $\eta'$.
The key observation here is that ${\cal L}^{(0)}$ contains the derivatives
of this field exclusively through the term
$\frac{1}{12}F^2(\partial \tpsi-D\theta)^2$. This implies that -- if the 
vacuum angle and the
singlet axial field are turned off -- only the vertices proportional to 
$\chi$ generate loops involving the propagation of an $\eta'$. 
Accordingly, the matching of the derivative terms 
is trivial: The coupling constants
$F, L_1,\,\ldots,\, L_5, L_9,L_{10}$ and $H_1$ are the same in the two 
versions of the theory. For $B,L_6,L_7,L_8$ and $H_2$, 
the matching conditions\footnote{Note that the low energy constants are
independent of the quark masses. 
The matching conditions do, therefore, not involve $M_\pi$, $M_K$, $M_\eta$,
but they do contain the mass scale set by the topological susceptibility,
which is related to the value of $M_{\eta'}$ in the chiral limit.} 
read \cite{Herrera 4}
\bea\label{SU3U3}
B^{\SU}\al=\al B\left\{1-\frac{2M_0^2}{3F^2}\lambda_0
+O(\nc^{-3})\right\}\no
L_6^{\SU}\al=\al L_6+ \frac{1}{72}\,
\left(\lambda_0-\frac{1}{32\pi^2}\right)+O(\nc^{-1})\no
L_7^{\SU}\al=\al-\frac{F^4(1+\Lambda_2)^2}{288\,\tau}+L_7+O(\nc^{-1})
\label{matching}\\
L_8^{\SU}\al=\al L_8- \frac{1}{12}\hspace{0.1em}\lambda_0+O(\nc^{-1})\no
H_2^{\SU}\al=\al H_2-\frac{1}{6}\hspace{0.1em}\lambda_0+O(\nc^{-1})\no 
\lambda_0\al =\al\lambda\,
\rule[-0.5em]{0.03em}{1.3em}_{\;\mu\rightarrow M_0}\co
\hspace{2em}
M_0^2= \frac{6\,\tau}{F^2(1+\Lambda_1)}\fs\nonumber\eea
The quantity $M_0$ is the mass of the $\eta'$ in the chiral limit.
This mass sets the scale of the logarithm contained in $\lambda_0$.
As a check, we note that in the expression for $M_0$,
the dependence of $\Lambda_1$ on the running scale of QCD 
cancels against the one of $\tau$, in
agreement with the fact that the mass of the $\eta'$ is renormalization group
invariant, also in the chiral limit (this is the reason for not expanding 
the denominator in that expression). 
The difference between $B^{\SU}$ and $B$ accounts for the
fact that the latter picks up a renormalization from the one loop graphs of
the effective theory, while the former does
not.
 
The difference between the values of $L_6$, $L_8$, $H_2$ in the two
versions of the effective theory is related to the difference between
the corresponding coefficients $\Gamma_6$, $\Gamma_8$, $\Delta_2$.
Indeed, one may check that in the matching relations
between the renormalized coupling constants, the divergences drop out.
The coupling constant $L_7$ is scale independent in both versions.
In the extended theory, $L_7$ is suppressed by the OZI-rule, while 
$L_7^{\SU}$ represents a term of order $\nc^2$.
The leading contribution arises from $\eta'$-exchange
and is inversely proportional to the square of the mass of the $\eta'$
 \cite{GL SU(3)}:
\bea L_7^{\SU}\simeq -\frac{F^2}{48 M_{\eta'}^2}\fs\nonumber\eea 
According to eq.~(\ref{matching}), the first order correction to 
this formula 
is determined by the OZI-violating coupling
constants $\Lambda_1$ and $\Lambda_2$: 
\bea \label{L7}L_7^{\SU}\al=\al-
\frac{F^2(1+\Lambda_2)^2}{48M_0^2(1+\Lambda_1)}+
L_7+O(\nc^{-1})\fs\eea
The term $L_7$ only enters as a correction of second order in $1/\nc$.
Note that the dependence on the running scale of QCD also cancels
out here: The relation (\ref{renLambda}) shows that the ratio 
$(1+\Lambda_2)^2/(1+\Lambda_1)$ is renormalization group
invariant. 

In the present section we considered those couplings of next-to-leading
order that survive if $\theta$ and $a_\mu^0$
are turned off -- the effective Lagrangian of ref.~\cite{GL SU(3)}
only accounts for these. As shown in appendix \ref{matching2}, the general 
expression of next-to-leading
order involves 11 additional terms. The matching relations for the
corresponding coupling constants are also given there. 

\setcounter{equation}{0}
\section{Anomalies}
\label{unnatural}
Above, we focussed on the natural parity part of the effective Lagrangian.
The photonic decays
$\pi^0\rightarrow\gamma\gamma$, $\eta\rightarrow\gamma\gamma$,
$\eta'\rightarrow\gamma\gamma$, for instance,  
are not covered, because these are contained 
in the unnatural parity part, which collects 
those terms that involve the tensor
$\epsilon^{\mu\nu\rho\sigma}$. We now extend the above discussion to these
and first consider the anomalies, which we analyze by means   
of the differential forms
\bea v\al=\al dx^\mu v_\mu\co\hspace{2em}
a=  dx^\mu a_\mu\co\hspace{2em} r=v+a\co\hspace{2em}l=v-a\co\hspace{2em}
d= dx^\mu\partial_\mu\fs
\nonumber\eea 
The quantities $dx^0,dx^1,dx^2,dx^3$ are treated as Grassmann variables. Their
product yields the standard volume element, 
$dx^\mu dx^\nu dx^\rho dx^\sigma=\epsilon^{\mu\nu\rho\sigma}d^4\!x$. 

The phase of the determinant of the Dirac operator is not invariant under
an infinitesimal U(3)$_{\indR}\times$U(3)$_{\indL}$ transformation of the
external fields, 
\bea V_{\indR}=1+i\,\alpha_{\indR}\co\hspace{2em} 
V_{\indL}=1+i\,\alpha_{\indL}\fs\nonumber\eea
The part of the phase change that depends on the gluon field
is compensated by the transformation $\delta\theta=\langle\alpha_{\indL}
-\alpha_{\indR}\rangle$ of the vacuum angle. The remainder is unique
only up to contact terms formed with the external fields $v_\mu$, $a_\mu$ and
$\theta$. In the standard convention, it is invariant under the
transformations generated by the vector charges, so that the change in the
effective action is of the form
\bea \al\al\delta S_{\eff}\{v,a,s,p,\theta\}= \!\int 
\langle(\alpha_{\indL}-\alpha_{\indR})\,\Omega\rangle\,.\nonumber\eea
An explicit formula\footnote{The sign of $\Omega$
is convention dependent; we use 
  the metric $\mbox{$+$$-$$-$$-$}$, set 
  $\epsilon_{\sss 0123} = +1$ and identify 
$\gamma_{\sss5}$ with 
  $\gamma_{\sss5} = -i 
  \gamma_{\sss 0}\gamma_{\sss 1}\gamma_{\sss 2}\gamma_{\sss 3}$.} 
for $\Omega$ was given in ref.~\cite{Bardeen}:
\bea  \label{Omega}
 \al\al\Omega=
\frac{\nc}{8\pi^2}\left\{F_v F_v+\mbox{$\frac{1}{3}$}D_v a D_v a 
+\mbox{$\frac{1}{3}$}\,i\,(F_va^2+4 a
  F_v a + a^2 F_v)+\mbox{$\frac{1}{3}$}a^4\right\}\,,\no
\al\al F_v= dv-iv^2\,,\hspace{2em}D_va=da-iva-iav\,.\eea
The expression for $\Omega$ is not unique, however -- in fact, the one given
is in conflict with the invariance of the effective action under the 
renormalization group. The problem arises, because the singlet
component of the axial external field transforms in a nontrivial manner under
this group. The differential form $a=dx^\mu a_\mu$ consists of two parts
(compare section \ref{dep theta}),
\bea\label{ab} a=\bar{a}+\mbox{$\frac{1}{6}$}D\theta\fs\eea
The first is renormalization group invariant and transforms as a gauge field
under chiral rotations. The second transforms
with $(D\theta)^{ren}=\za^{-1}D\theta$ under the renormalization group, 
but is invariant
under U(3)$_{\indR}\times$U(3)$_{\indL}$. Inserting the decomposition in the
expression for $\Omega$, we obtain a sum of three terms,
\bea \Omega=\Omega_0+\Omega_1+\Omega_2\fs\nonumber\eea
The first is obtained from $\Omega$ by replacing $a$ with $\bar{a}$:
\bea\label{Omega0} 
\hspace*{-2em}\Omega_0=
\frac{\nc}{8\pi^2}\left\{F_v F_v+\mbox{$\frac{1}{3}$}D_v \bar{a} D_v \bar{a} 
+\mbox{$\frac{1}{3}$}\,i\,(F_v\bar{a}^2+4 \bar{a}
  F_v \bar{a} + \bar{a}^2 F_v)+\mbox{$\frac{1}{3}$}\bar{a}^4\right\}\co\eea
where $D_v\bar{a}=d\bar{a}-i v \bar{a}-i \bar{a} v$. It is
renormalization group invariant. The remaining two terms
are given by
\bea
\Omega_1=\frac{\nc}{36\pi^2}\,\{D_v\bar{a}\,\langle da\rangle-
i(F_v\bar{a}-\bar{a}F_v) D\theta\}\co\hspace{2em}
\Omega_2=\frac{\nc}{216\pi^2}\,\langle da\rangle\,\langle da\rangle
\co\eea
and transform with $\Omega_1^{ren}=\za^{-1}\Omega_1$, 
$\Omega_2^{ren}=\za^{-2}\Omega_2$.
If we were to identify the anomaly of the effective action 
with $\Omega$, this functional would fail to be independent of the running
scale of QCD, in contradiction with eq.~(\ref{rg}).

The problem is readily solved: The extra terms $\Omega_1$ and $\Omega_2$
represent the anomalies generated by two contact terms,
\bea P_1=\frac{\nc}{36\pi^2}\,\langle\bar{a}D_v\bar{a}\rangle 
D\theta\co\hspace{2em}
P_2=\frac{\nc}{216\pi^2}\,\langle a\rangle \langle da\rangle d\theta
\fs\eea
Removing these, the anomaly takes the renormalization group invariant
form 
\bea\label{anomaly Seff} \delta S_\eff\{v,a,s,p,\theta\}=
\int\!\langle (\alpha_{\indL}-\alpha_{\indR})\,\Omega_0
\rangle\fs
\eea
Since the contact terms $P_1$, $P_2$ vanish for
$\langle a_\mu\rangle=\partial_\mu\theta=0$, they only matter
when considering correlation functions that contain the operators $A_\mu^0$,
$\omega$.

\setcounter{equation}{0}
\section{Wess-Zumino-Witten term}\label{WZW}
Within the effective theory, the anomalies are accounted for by
the Wess-Zumino-Witten term. The standard expression for this term
reads \cite{WZW}
\bea\label{SWZW} S_{\mbox{\tiny{WZW}}}\{U,v,a\}\al=\al
-\frac{i\,\nc}{240\pi^2}\int_{M_5}\!\!\!\langle
\Sigma^{\;5}\rangle-\frac{i\,\nc}{48\pi^2} \int_{M_4}\!\!\!
\left\{W(U,r,l)-W({\bf 1}, r,l)\right\}\co \no
W(U,r,l)\al=\al \langle U\,l^{\,3}U^\dagger r+\mbox{$\frac{1}{4}$}\,U\,l\,
U^\dagger r\,
U\,l\,U^\dagger r
   +i\,Udl\,l\,U^\dagger r + i\, dr\,U\, l\,U^\dagger r \nonumber\\\al\al
-i\,\Sigma\, l\, U^\dagger r \,U\, l
+\Sigma\, U^\dagger dr\, U\, l-\Sigma^{2} U^\dagger r\, U\, l+
\Sigma\, l\,dl+
\Sigma\, dl\,l\nonumber\\\al\al -i\,\Sigma\,
l^{\,3} +\mbox{$\frac{1}{2}$}\,\Sigma\, l\,\Sigma\,l
-i\, \Sigma^{3} \, l \rangle -\RL
\co\eea
with $U\in\mbox{U(3)}$ and $\Sigma\equiv U^\dagger dU$. 
The first term is an integral 
over a field $U(x,x^5)$ that smoothly interpolates between 
$U(x,0)={\bf 1}$ and $U(x,1)=U(x)$. The Grassmann algebra is supplemented with
a fifth element $dx^5$ and the integration extends over the five dimensional
manifold $M_5$, which represents the direct product of Minkowski space with
the interval $0<x^5<1$. The integral is independent of the particular
interpolation chosen to connect $U(x)$ with the unit matrix.
In the second term, the integration only extends over 
Minkowski space, $M_4$. The operation 
$\RL$ requires an interchange of the 1-forms
$r$ and $l$ as well as an interchange of $U$ and $U^\dagger$. By construction,
an infinitesimal chiral rotation of the variables $U,v$ and $a$ generates
the change
\bea \delta S_{\mbox{\tiny{WZW}}}\{U,v,a\}=\int\!\langle 
(\alpha_{\indL}-\alpha_{\indR})\,\Omega\rangle\co
\eea
where $\Omega$ is the 4-form specified in eq.~(\ref{Omega}). 

For the reason given in the preceding section, we remove the contact terms 
$P_1$, $P_2$ and define the Wess-Zumino-Witten part of the effective Lagrangian
through
\bea \int\!\!dx\,{\cal L}_{\mbox{\tiny WZW}}\equiv
S_{\mbox{\tiny{WZW}}}\{U,v,a\}
-\int\!(P_1+P_2)\fs\eea
This ensures that the term ${\cal L}_{\mbox{\tiny WZW}}$
accounts for the anomalies of 
QCD in the renormalization group invariant form (\ref{anomaly Seff}):
\bea \delta\!\!\int\!\!dx\,{\cal L}_{\mbox{\tiny WZW}}=
\int\!\langle (\alpha_{\indL}-\alpha_{\indR})\,\Omega_0
\rangle\fs\eea 
Note, however, that the term ${\cal 
  L}_{\mbox{\tiny WZW}}$ as such is not renormalization group invariant. 
Since the scale dependent pieces contained therein  are 
gauge invariant (see appendix 
\ref{decomposition}), we could remove these and arrive at a scale 
invariant expression. The drawback of such a choice
is that it interferes with the large $\nc$ counting rules: As demonstrated
in the next section, our definition of ${\cal L}_{\mbox{\tiny WZW}}$ 
is singled out by the property that it represents the 
leading unnatural parity part of the effective Lagrangian. 

\setcounter{equation}{0}
\section{Unnatural parity part beyond leading order}
\label{unnatural ho}
We denote the unnatural parity part of the effective Lagrangian by
$\tilde{\cal L}_\eff$. As we did not find a complete list for the terms 
of order $p^4$ in the literature, we briefly outline the 
construction.

Once the WZW-term is removed, the unnatural parity part 
also becomes gauge invariant under U(3)$_{\indR}\times$U(3)$_{\indL}$. It is 
convenient to
express the Lagrangian in terms of the variables 
$U$, $\tpsi$, $v_\mu$, $a_\mu$, $D_\mu\theta$ and their derivatives.
Terms involving derivatives of $\tpsi$ can be integrated by parts. 
At order $p^4$, charge conjugation invariance then
allows six independent invariants ($\tilde{F}^{\mu\nu}\equiv\frac{1}{2}
\epsilon^{\mu\nu\rho\sigma} F_{\rho\sigma}$):
\begin{eqnarray*}
\al\al\tilde{\cal L}_{\eff}={\cal L}_{\mbox{\tiny WZW}}+
\tilde{V}_1\,  i \br \tilde{R}^{\mu \nu } D_\mu U D_\nu
U^\dega +  \tilde{L}^{\mu \nu } D_\mu U^\dega D_\nu U \ke
+\tilde{V}_2\, \br \tilde{R}^{\mu \nu } U L_{\mu \nu }U^\dega \ke\no \al\al
\hspace{6em}
+ \tilde{V}_3\, \br \tilde{R}^{\mu \nu } R_{\mu \nu } +\tilde{L}^{\mu \nu }
L_{\mu \nu }   \ke  
+  \tilde{V}_4\,   i D_\mu \theta\,  \br
  \tilde{R}^{\mu \nu } D_\nu U U^\dega -\tilde{L}^{\mu \nu } U^\dega  D_\nu
  U\ke  
\no\al\al\hspace{6em} +\tilde{V}_5
\,( \br \tilde{R}^{\mu \nu } \ke 
\br  R_{\mu \nu } \ke+  \br\tilde{L}^{\mu \nu }
\ke \br L_{\mu \nu } \ke) 
+\tilde{V}_6\, \br \tilde{R}^{\mu \nu } \ke \br  L_{\mu \nu } \ke+O(p^6)\fs
\end{eqnarray*}
On account of parity, all of the potentials are odd functions of 
$\tpsi$, except for $\tilde{V}_4$, which is even.

In the $1/\nc$ expansion, the leading contribution to the potentials
$\tilde{V}_1$, $\tilde{V_2}$ and $\tilde{V_3}$ is 
linear in $\tpsi$ and contains a coupling constant of $O(1)$, while 
$\tilde{V}_4$ reduces to a constant that is also of this order.
Since the remaining two terms involve two flavour traces, their expansion  
starts at order $1/\nc$. This implies that the simultaneous expansion in 
powers of $p$ and $1/\nc$ is dominated by the WZW-term,
which represents a contributions of order $\nc\,p^4=O(\delta)$. The remainder
is of order $\delta^2$ or higher:  
\bea\label{Ltilde} \tilde{\cal L}_{\eff}= {\cal L}_{\mbox{\tiny WZW}}+
\tilde{\cal L}^{(2)}+\tilde{\cal L}^{(3)}+\ldots\eea

The general expression for the next-to-leading order Lagrangian contains
contributions of the type $p^4$ and $\nc\, p^6$. The former can be extracted 
from the representation given above:
\bea\al\al\hspace{-2em}\tilde{\cal L}^{(2)}_{p^4}=
\tilde{L}_1\,i\hspace{0.1 em}\tpsi\,   \br \tilde{R}^{\mu \nu } D_\mu U D_\nu
U^\dega +  \tilde{L}^{\mu\nu} D_\mu U^\dega D_\nu U \ke 
+\tilde{L}_2\,\tpsi\, \br \tilde{R}^{\mu\nu} U L_{\mu \nu }U^\dega \ke\\ 
\al\al+ \tilde{L}_3\,\tpsi\, \br \tilde{R}^{\mu \nu } R_{\mu \nu  }
+ \tilde{L}^{\mu \nu } L_{\mu \nu  } \ke 
+  \tilde{L}_4\,   i D_\mu \theta\,  \br
 \tilde{R}^{\mu \nu} D_\nu U U^\dega - \tilde{L}^{\mu \nu }U^\dega  D_\nu U\ke 
\fs \non \eea
At order $\nc\, p^6$, many invariants can be formed, 
in particular also
terms proportional to the quark mass matrix. Below we will discuss only a 
selection thereof: The terms relevant in connection with the radiative
transitions. 

Note that ${\cal L}_{\mbox{\tiny WZW}}$ is not independent of the scale used
to renormalize the singlet axial current.
The scale dependent part is gauge invariant, but represents a 
contribution of leading order in the $1/\nc$ expansion
and must be retained for the relation (\ref{Ltilde}) to hold.
As in the case of the natural parity part, renormalization
group invariance thus requires specific contributions of nonleading order.
The phenomenon arises from the fact that 
the loop graphs responsible for the anomalous dimension of the singlet 
current violate the OZI rule, which implies that some of the effective 
coupling constants contained in $\tilde{\cal L}^{(2)}$ 
must compensate for the scale dependence of ${\cal L}_{\mbox{\tiny WZW}}$.
The relevant terms are those of order $p^4$, all of which were listed above. 
The dependence of the coupling constants on the running scale
of QCD follows from the decomposition of the WZW-term
given in appendix \ref{decomposition}: The sum ${\cal L}_{\mbox{\tiny WZW}}+ 
\tilde{\cal L}^{(2)}_{p^4}$ is 
renormalization group invariant to order $\delta^2$, provided 
$\tilde{L}_1,\ldots,\,\tilde{L}_4$ are renormalized according
to
\bea \tilde{L}_1^{ren}\al=\al\za \tilde{L}_1 -\kappa\co\hspace{1em}
\tilde{L}_2^{ren}=\za \tilde{L}_2 -\kappa\co\hspace{1em}
\tilde{L}_3^{ren}=\za \tilde{L}_3 -\kappa\co\no
\tilde{L}_4^{ren}\al=\al \za \tilde{L}_4 + \kappa \co\hspace{1em}
\kappa= \frac{\nc\,(\za-1)}{144\pi^2}\fs\eea
The quantity $\kappa$ is of $O(1)$, like the coupling constants
themselves. 

\setcounter{equation}{0}
\section{Radiative transitions}\label{transitions}
As an illustration, we consider the radiative transitions 
$\pi^0\rightarrow\gamma\gamma$, $\eta\rightarrow\gamma\gamma$,
$\eta'\rightarrow\gamma\gamma$ \cite{Bijnens}. 
At leading order, the corresponding part of
the effective Lagrangian is obtained from the Wess-Zumino-Witten term with 
\bea r=l=-e\, Q  A\co\nonumber\eea 
where $Q=\mbox{diag}\{\frac{2}{3},-\frac{1}{3},-\frac{1}{3}\}$ represents 
the charge matrix of the light quarks and $A=dx^\mu A_\mu$ is the 1-form 
associated with the electromagnetic field. 
Setting $U=e^{i\phi}$, the terms linear in $\phi$ and 
quadratic in $A$ reduce to
\bea
S_{\mbox{\tiny{WZW}}}\{U,v,a\}=\frac{\nc\,e^2}{8 \pi^2}\int\!
\langle Q^2 d\phi\rangle A\, dA=-\frac{\nc\,\alpha}{4 \pi}\int\!\!d^4\!x
\langle Q^2 \phi\rangle F_{\mu\nu}\tilde{F}^{\mu\nu}\fs\nonumber\eea
Concerning $\tilde{\cal L}^{(2)}_{p^4}$, only the combination
$\tilde{L}_2+2\tilde{L}_3$ of coupling constants matters for the photonic 
transition 
matrix elements. The net effect of the first order OZI violations
is that the trace $\langle Q^2\phi\rangle$ appearing in 
the WZW-term is replaced by
\bea \langle Q^2\phi\rangle\rightarrow
\langle Q^2 \phi\rangle +\mbox{$\frac{1}{3}$}
K_1\langle Q^2\rangle\langle \phi\rangle \co\nonumber\eea
with\footnote{in ref. \cite{Montpellier} this coupling constant is denoted by
  $\Lambda_3 =K_1$.}
$K_1=-48\pi^2(\tilde{L}_2+2\tilde{L}_3)/\nc$. 
The scaling laws for the coupling constants $\tilde{L}_2$, $\tilde{L}_3$ 
imply that the renormalization of $K_1$ is of the same form as
the one for $\Lambda_2$: 
\bea\label{renK} 1+K_1^{ren}=\za(1+K_1)\fs\eea 
One readily checks that this indeed compensates the 
renormalization of the singlet field 
in eq.~(\ref{renpsi}), so that the result for the transition amplitude is
renormalization group invariant.

The Lagrangian $\tilde{\cal L}^{(2)}$ contains two further categories of 
contributions: Terms of $O(\nc)$ with 6 derivatives and chiral
symmetry breaking effects of $O(\nc\hspace{0.1em} m\hspace{0.1em} p^4)$. We
denote these by 
$\tilde{\cal  L}^{(2)}_{p^6}$ and $\tilde{\cal  L}^{(2)}_{\chi}$, respectively.
As far as the photonic transitions are concerned, the first contains 
two independent contributions, which may be written in the form:
\bea \tilde{\cal L}^{(2)}_{p^6}=\tilde{L}_5\, e^2 \langle Q^2\phi\rangle 
F_{\mu\nu}\,\wave\tilde{F}^{\mu\nu}+\tilde{L}_6\,
e^2\langle Q^2\wave\phi\rangle  F_{\mu\nu}
\tilde{F}^{\mu\nu}\fs\nonumber\eea
Using the equations of motion, the second term can be absorbed in
the couplings occurring in $\tilde{\cal L}^{(2)}_{p^4}$ and 
$\tilde{\cal  L}^{(2)}_{\chi}$. 
Since the term with $\wave F^{\mu\nu}$ only matters for off-shell photons, 
we can ignore this part of the Lagrangian altogether.

Finally, we consider the chiral symmetry breaking terms. 
As only neutral mesons
are involved, the matrices $\phi,\,Q$ and $\chi=2B\hspace{0.05em}m$ commute,
so that there 
is 
only one independent invariant with a single flavour trace. We again extract
a normalization factor and denote the coupling constant by $K_2$:
\bea \tilde{\cal L}^{(2)}_\chi=- \frac{ \alpha \nc\, K_2 }{4 \pi}
     \langle Q^2 \chi \phi \rangle F_{\mu\nu}
\tilde{F}^{\mu\nu}\fs\nonumber\eea
The net result for the effective Lagrangian that describes the photonic decays
to first nonleading order thus reads:
\bea {\cal L}_{P\rightarrow\gamma\gamma}=-\frac{ \alpha \nc}{4 \pi}\left\{
\langle Q^2 \phi\rangle +\mbox{$\frac{1}{3}$}
K_1\langle Q^2\rangle\langle \phi\rangle 
+ K_2 \langle Q^2 \chi \phi \rangle \right\}
F_{\mu\nu}\tilde{F}^{\mu\nu}\fs\eea
The coupling constant $K_1$ describes the corrections generated by the
violations of the OZI rule and $K_2$ accounts for the breaking of
chiral symmetry. While $K_1$ represents a term of order $1/\nc$ and depends 
on the running scale of QCD according to eq.~(\ref{renK}), the constant $K_2$ 
is of order 1 and is renormalization group invariant.

\setcounter{equation}{0}
\section{Summary and conclusion}

The effective theory of QCD with three colours is well known.
In that framework, the effective Lagrangian consists of 
all terms respecting chiral symmetry. The chiral perturbation series
amounts to an expansion in powers of momenta and
quark masses. In the present paper, we have examined the extension of this
effective theory required
to analyze the low energy properties of QCD 
in the limit where the number of colours, $\nc$, is treated as large.  
In that case, the situation is more intricate, because there is an
additional low energy scale, related to the 
mass of the $\eta'$. As discussed in detail, the standard expansion
in powers of momenta and quark masses, which we refer to as the $p$-expansion,
cannot be used when $\nc$ becomes large: In order to coherently analyze the
  low energy properties of the $\eta'$, the two quantities appearing in the 
$\eta'$-propagator -- the mass and the square of the momentum -- 
must be treated on equal footing. 
We exploit the fact that $M_{\eta'}^2$ represents a term of
$O(1/\nc)$, so that an expansion that counts $p^2$, $m$ and $1/\nc$ 
as quantities of the same order does have the desired property.
We refer to the corresponding low energy
expansion as the $\delta$-expansion. It orders the triple series in 
$p=O(\sqrt{\delta})$, $m=O(\delta)$ and $1/\nc=O(\delta)$ 
by collecting terms that are of the same
order in $\delta$.

In the limit $\nc \rightarrow \infty$, 
$g^2 \nc$ fixed, the perturbative analysis of QCD determines the order in 
$1/\nc$ of the various correlation
functions, according to eq.~(\ref{countG}). These counting rules only hold
for nonexceptional momenta. In particular, the low energy singularities 
generated by the Goldstone bosons (including the $\eta'$) give rise to
contributions that violate these rules. We have shown, however, that
the generating functional $S_{\eff}
\{v,a,s,p,\theta\}$ does admit a coherent expansion in powers of $\delta$.

What we then discussed in some detail are the implications of this
property of the generating functional for the effective
Lagrangian. The reason why this step is not straightforward is 
closely related to the fact that the dynamical variables of the effective 
theory do not have physical meaning -- their choice is inherently ambiguous. 
This entails that statements about the effective
Lagrangian can only be true modulo a change in the dynamical variables. What
we were able to show is that there exists a class of such variables, for which
the coupling constants in the effective Lagrangian are at most of $O(\nc)$.
This is natural, because the correlation
functions collected in the generating functional obey the same bound. 
Furthermore, we have shown that, like for the generating functional, 
the dependence on the vacuum angle $\theta$ is
suppressed in the large $\nc$ limit: At order $\delta^n$, the effective 
Lagrangian is a polynomial in $\theta$, of degree $n+2$.   

The outcome of our investigation boils down to a remarkably simple
construction recipe for the effective Lagrangian that holds to any  
given order in $\delta$:
\begin{description}
\item[(i)] 
\hspace*{0.8em}Apart from the WZW-term, the Lagrangian is manifestly invariant under local
U(3)$_{\indR}\times$U(3)$_{\indL}$ transformations.

\item[(ii)] 
\hspace*{0.5em}At order $\delta^n$, the Lagrangian represents a polynomial formed with the 
fields $U$, $U^\dagger$, $s$, $p$, $R_{\mu\nu}$, $L_{\mu\nu}$,  $\tpsi$, 
$D_\mu\theta$ and their covariant derivatives. The expression 
contains terms that are at most of
order $p^{2n+2}$. 
 
\item[(iii)] The coupling constant associated with a term in the Lagrangian is
  of order $\nc^{2-k}$,
where $k$ counts the number of traces plus the number of factors with $\tpsi$,
$D_\mu \theta$ or derivatives thereof (note that $k \ge 1$).
\end{description}
In the text, we explicitly give the expressions for the
Lagrangians of order
$1$ and order $\delta$, but list only a selection of the vertices occurring at
next-to-next-to leading order, for practical reasons: We expect 
the full Lagrangian of order $\delta^2$ to contain about 100 terms. 

We have also shown that the effective theory is consistent with
periodicity in the vacuum angle 
$\theta$. This is not evident, a priori, because the effective Lagrangian
truncated at a given order in 
$\delta$ is actually a polynomial in $\theta$. The paradox is resolved by
observing that the periodicity is not a necessary feature of the Lagrangian
itself -- only the corresponding effective 
action needs to be periodic, and this is the case.

A corollary of our results concerning the structure of the effective 
Lagrangian is that the transformation of Kaplan and 
Manohar is forbidden at large $\nc$: For $\theta \neq 0$, this 
transformation involves the vacuum angle through a factor
$e^{-i\theta}$ and thus generates a modification of the effective Lagrangian
with a nonpolynomial dependence on $\theta$. This is in
conflict with the properties of the effective theory at large $\nc$.

As is well known, the singlet axial current
$A_\mu^0$ carries anomalous dimension and thus depends on the running scale of
QCD. Moreover, the renormalization group mixes the operators $\omega$ 
and $\partial^\mu \!A^0_\mu$. In the effective action, the corresponding 
external fields are the trace $\langle a_\mu\rangle$ and the vacuum angle
$\theta$. We have shown that
their renormalization group properties follow
from symmetry considerations alone: While the vacuum angle is scale
independent, the singlet field $\br a_\mu \ke $ transforms
inhomogeneously under scale transformations and picks up a contribution
proportional to the gradient of the vacuum angle.
We have worked out the consequences for the dynamical variables
and coupling constants of the effective theory. In the leading order
Lagrangian ${\cal L}^{(0)}$, the coupling constant $F$ is scale
independent, while $B$ and $\tau$ are multiplicatively renormalized.
Some of the fields contained therein, however, transform
inhomogeneously, so that ${\cal L}^{(0)}$ does not remain invariant when the
running scale is varied. The origin of the problem -- the
anomalous dimension of the singlet axial current --
is due to graphs of nonleading order. This implies that the change in 
${\cal L}^{(0)}$ produced by a change of scale is an effect of order 
$1/\nc$ and is eaten up by a suitable shift of the couplings 
occurring in ${\cal L}^{(1)}$, so that the effective action does remain 
invariant. Quite generally, the action of the renormalization 
group on the effective coupling constants occurring at a given order
mixes these with the lower order couplings. Physical quantities only
involve scale independent combinations of coupling constants, which are
easily identified.

In the remainder of the paper, we have extended the analysis to the 
unnatural parity part of the Lagrangian and have shown that
the scale independence of the effective theory can be made manifest also in
this sector. At leading order, the unnatural parity part is given by
the Wess-Zumino-Witten term, which accounts for the anomalies within the
effective theory. The straightforward extension of this
term to the case of U(3)$_\indR\times$U(3)$_\indL$, however, fails to be 
invariant under
the renormalization group. In part, the deficiency only concerns contact
contributions that are readily removed.
The remainder still contains scale dependent contributions, but
these are gauge invariant, so that the modification produced by a change of
the running scale may be absorbed in the coupling constants occurring at
the next order of the expansion. The relevant terms are those 
of unnatural parity at order $\delta^2$. We could instead have modified the
expression for ${\cal L}_{\mbox{\tiny{WZW}}}$, by adding suitable gauge
invariant 
pieces that make it scale independent. This, however, would upset the 
large $\nc$ counting rules, so that the leading contribution in the 
$\delta$-expansion of the unnatural parity  
Lagrangian would then not coincide with ${\cal L}_{\mbox{\tiny{WZW}}}$.

To establish contact with the standard low energy theory of QCD, we have given
the 
explicit matching relations between the low energy constants relevant at large
and at fixed $\nc$. We have also discussed the extension needed to investigate
singlet currents in the standard framework. Some features, such as the
dependence of the effective coupling constants on the running
scale of QCD apply to
both versions of the theory. Others do not: The Kaplan-Manohar transformation,
for instance, is in conflict with the properties of QCD only if the number of
colours is treated as large -- at fixed $\nc$, the effective theory is 
invariant under this operation.

\appendix
\renewcommand{\theequation}{\thesection.\arabic{equation}}
\setcounter{equation}{0}
\section{Construction of the effective Lagrangian}
\label{appendix A}
In the present appendix, we show how the reasoning of sections
\ref{simultaneous expansion}--\ref{chiral symmetry} can be extended to
construct the full effective Lagrangian. As input, we use the large
$\nc$ properties of the Green functions of QCD. We first switch the  
quark masses off and use
the large $\nc$ counting rules for the scattering 
amplitudes of the
pseudoscalar mesons. These allow us to establish corresponding counting rules
for the interaction vertices of the effective Lagrangian. Then, we
generalize the argument to those vertices that describe the response of the
system to the perturbations generated by the external fields
and finally discuss the consequences of the Ward identities of chiral symmetry.

\subsubsection*{Scattering amplitude at large \boldm{$\nc$}}
If the number of colours is sent to infinity and the quark masses are turned
off, the spectrum of QCD contains nine massless pseudoscalar mesons. 
The scattering amplitudes describing the interaction
of any number of these particles in the initial and final state can be
extracted from the connected correlation functions formed with
the corresponding number of axial currents, using these currents as 
interpolating fields. The correlation functions represent quantities of order
$\nc$, irrespective of the number of currents contained therein.
In view of the fact that the one particle matrix elements 
of the currents, $\lvac\!A^a_\mu|\pi^ b\rangle$, are of order $\sqrt{\nc}$,
the scattering amplitude for $n=n_{i}+n_{f}$ mesons is at most
of order $\nc^{1-n/2}$.

The scattering amplitude contains singularities in the low energy region.
In particular, for $n\geq 6$,
it contains one particle reducible contributions, 
describing a sequence of collisions, connected by the exchange of single
particles. Denoting the
number of exchanged particles by $\ell$, there are $\ell+1$ irreducible 
parts. The
number of meson lines entering or leaving the irreducible parts adds up
to $n+2\ell$. Applying the counting rule to the irreducible
parts, the resulting contribution to the scattering amplitude represents
a term of order $\nc^k$, with $k=(\ell+1)-\frac{1}{2}\,(n+2\ell)=
1-\frac{1}{2}\,n$.
This shows that the singularities generated by one particle exchange manifest
themselves already at leading order.

Unitarity relates the imaginary part to the square of the scattering
amplitude. The relation implies that this amplitude contains
further singularities. The exchange of a pair of mesons (two particle
intermediate states in the unitarity relation), for instance, generates a
branch cut in the one-particle-irreducible parts.  
The contribution from the two particle cut, however,
only shows up at order $\nc^{-n/2}$: If the two parts connected by the two
exchanged particles involve $n_1$ and $n_2$ mesons, respectively, we have
$n_1+n_2=n+4$, so that the overall power of $\nc$ is given by
$(1-\frac{n_1}{2})+(1-\frac{n_2}{2})=-\frac{n}{2}$. Exchanges of more than two
particles between the 
same irreducible parts are
suppressed even more strongly. This means that the branch cuts required
by unitarity only show up at nonleading orders of the $1/\nc$ expansion.
The leading order contributions only contain those singularities
that arise from one-particle-exchange. Moreover, at leading order, 
the one-particle-irreducible parts reduce to polynomials  
of the momenta.

At leading order of the $1/\nc$ expansion, the structure of the 
scattering amplitude is the same as the one of the tree graphs of a
pseudoscalar field theory. 
We identify the dynamical variables with the dimensionless
fields $\phi^0(x)$, $\ldots$ , $\phi^8(x)$ introduced
in eq.~(\ref{Upi}) 
and represent the interaction Lagrangian in the symbolic form
\bea\label{LN} {\cal L}_{\eff}= \sum_{k,\,n} 
g(k,n)\times\partial^{\,k}\!\times\phi^n\co\eea
where the flavour and Lorentz
structure of the vertices is suppressed.  The integers $k$ and $n$ 
merely count the number of derivatives and fields occurring in the vertex in 
question and
$g(k,\,n)$ represents the corresponding effective coupling constant
-- in general, there are several, 
independent vertices of the same symbolic structure. Lorentz invariance 
implies that $k$ is even.

The terms quadratic in $\phi(x)$ are given in eqs.~(\ref{LU1}) and (\ref{Lpi}).
The corresponding coupling constants are $g(0,2)\sim \tau_{\mbox{\tiny
GD}}=O(1)$ and $g(2,2)\sim F^2=O(\nc)$. For $n>2$, the  
coupling constant $g(k,n)$ generates a tree graph contribution to
the one particle irreducible scattering amplitude with $n$ mesons. The
contribution is of the symbolic form
$g(k,n)\, p^k\,F^{-n}$, where $p$ stands for the momenta of the
particles. The comparison with the counting rule for the scattering amplitude
suggests that the coupling constant $g(k,n)$ can at most be of 
order $\nc$.

\subsubsection*{Freedom in the choice of the dynamical variables}
Actually, the argument just given runs in the wrong direction: It 
only shows that if the Lagrangian exclusively
contains vertices of order $\nc$, then the
corresponding scattering amplitude does
obey the large $\nc$ counting rule -- the converse is not true. A counter
example can be constructed as follows.
As discussed in section \ref{effective
  theory}, the effective Lagrangian is not unique, because its form depends
on the choice of variables. We may for instance subject the field $U(x)$ to
the transformation $U'=U\exp i f(\psi+\theta)$. The operation 
preserves the transformation law (\ref{trafopsi}), irrespective of the choice
of the function $f(x)$. We may choose one that grows with $\nc$. Suppose that
the coupling constants are of order $\nc$ and express the Lagrangian in terms
of the new variables. The resulting expression describes the same physics,
but contains effective coupling constants that grow more rapidly 
than with the first power if $\nc$ becomes large.

The freedom in the
choice of the dynamical variables is related to the fact that, in 
the scattering amplitude, all of the momenta are on the mass shell. 
It is well known that different vertices may give rise to the same on-shell
matrix elements. The requirement that the on-shell matrix elements of a
given set of vertices reproduces certain contributions occurring in 
the scattering amplitude only fixes these matrix elements up to terms
that vanish on the mass shell of the colliding particles, $p_i^2=M_i^2$.
In the chiral limit, eight of these are massless, $M_i=0$, while the
mass of the ninth is given by $M_{\eta'}$.   
The off-shell extension involves an ambiguity of the form 
$\sum_i (p_i^2-M_i^2)\,c_i(p_1,\ldots\,p_n)$. 
In coordinate space, this ambiguity 
corresponds to terms in the Lagrangian that are proportional to the 
equation of motion, which at leading order in the $\delta$-expansion
of the massless theory
is of the form $\wave\, \phi+\bar{\tau}
\langle\phi\rangle=h(\phi)$, with  
$\bar{\tau}=2\tau/F^2$. The right hand side, $h(\phi)$, 
consists of a series of terms that contain three or more
fields. 

This observation may be used to determine the ambiguity
in the effective Lagrangian in an iterative manner. Suppose that
the tree graphs of ${\cal L}_{\eff}$ and ${\cal L}_{\eff}^\prime$ 
generate the same on-shell scattering matrix elements.
As discussed in section \ref{effective theory}, we may 
choose the variables such that the terms that are quadratic in the 
meson fields
are the same, so that the difference $\Delta{\cal L}={\cal L}_{\eff}'-{\cal
  L}_{\eff}$ 
only contains vertices with four or more meson fields. Suppose now that the
four-particle scattering amplitudes generated by these Lagrangians 
coincide. This property implies that, up to a total
derivative, the terms of order $\phi^4$ contained in $\Delta{\cal L}$ 
can be written in the form 
$\sum_k\,c(k)\times (\wave \phi+\bar{\tau} \langle\phi\rangle)
\times\partial^k\times\phi^3$. We may replace 
$\wave \phi+\bar{\tau} \langle\phi\rangle$ with 
$\wave \phi+\bar{\tau} \langle\phi\rangle-h(\phi)$, because the
extra contributions contain six or more meson fields. It therefore suffices
to transform the variables in ${\cal L}_{\eff}'$ with
$\phi\rightarrow \phi+F^{-2} \sum_k\,c(k)\times\partial^k\times\phi^3$: 
At order $\phi^4$, the operation reduces the difference between the two
Lagrangians to a total derivative. Since a change of variables does not modify
the physics, the new version of ${\cal L}_{\eff}^\prime$ yields the same
scattering matrix elements as the original one, irrespective of the number of
particles participating in the collision. Iterating the procedure, we may
extend this analysis to terms with an arbitrary number of fields. 
We conclude that the
on-shell scattering matrix elements unambiguously determine the effective 
Lagrangian, except for
two inherent degrees of freedom: Choice of the dynamical variables
and total derivatives. The first reflects the ambiguities occurring in the
extension off the mass shell, the second concerns the extension off the 
energy-momentum shell.

We will fix the choice of variables when specifying the explicit expressions
for the first few terms of the derivative expansion. For the moment,
we only exploit the fact that there is a set of coupling constants
$g(k,n)=O(\nc)$, for which the tree graphs of the effective Lagrangian do
reproduce the scattering amplitudes at leading order of the $1/\nc$ expansion. 
This only excludes those transformations of variables that generate coupling 
constants growing more rapidly than with the first
power of $\nc$.

\subsubsection*{External fields}

Let us now turn on the external fields. The above analysis is readily extended
to this case. We again consider the massless theory and use the symbol
$j_i=\qbar\, \Gamma_i q$ 
to denote any one of the quark currents. The large $\nc$ counting rules
for the correlation functions of these operators 
were given in section \ref{gluodynamics}. We first generalize this to matrix
elements between asymptotic states and consider  
\bea
G_{n_j\,n_\omega\,n_i\,n_f}=
\langle f\,|\, T\,j_1(x_1)\cdots j_{n_j}(x_{n_j})\,
\omega(y_1)\cdots\omega(y_{n_\omega})
|\,i\,\rangle_c\co\nonumber\eea
where $|\,i\,\rangle$ and $\langle \,f\,|$ represent states with $n_i$ 
incoming and $n_f$ outgoing mesons, respectively. The behaviour
of $G_{n_j\,n_\omega\,n_i\,n_f}$ at large $\nc$ is established
in the same manner as for the scattering matrix: The matrix element
is related to the residue of the poles occurring in a correlation function
that, in addition to the operators listed, involves $n=n_i+n_f$ axial
currents, which play the role of the interpolating fields, while the 
operators $j_1(x_1)\cdots \omega(y_{n_\omega})$ are treated as spectators.
Denoting the contribution to $G_{n_j\,n_\omega\,n_i\,n_f}$ that arises from
graphs with $\ell$ quark loops by $G_{n_j\,n_\omega\,n_i\,n_f}^\ell$,
the generalization of eq.~(\ref{countG}) reads
\bea\label{countGn}G_{n_j\,n_\omega\,n_i\,n_f}^\ell=
O\left(\nc^{2-\ell-n_\omega-\frac{1}{2}n_i-\frac{1}{2}n_f}\right)
\co\hspace{2em}\ell=1,2,\ldots
\eea
We may also check that, like in the case of the scattering amplitude, 
the singularities generated by
one particle exchange manifest themselves at leading order, while 
the unitarity cuts only appear at nonleading orders. In the large $\nc$ limit,
the one particle irreducible parts can therefore again be expanded in the
momenta.

The above counting rule shows that the matrix elements of all of the quark 
currents behave in the 
same manner in the large $\nc$ limit, while those of the
operator $\omega(x)$ are suppressed. For the counting of powers relevant
at low energies, on the other hand, the external field $\theta(x)$ counts as 
a term of order 1, $v_\mu(x), a_\mu(x)=O(\sqrt{\delta})$ and
$s(x),p(x)=O(\delta)$. In the following bookkeeping, we 
do not distinguish between $v_\mu(x)$ and $a_\mu(x)$, nor between
$s(x)$ and $p(x)$ and write
the effective Lagrangian in the symbolic form
\bea\label{Lsymb} 
{\cal L}_{\eff}\al=\al\sum e(k,\,k_\theta)\times\nc^{2-k_\theta}
\times\partial^{\,k}\times\theta^{k_\theta}\\
\al+\al
\sum g(k,\,n,\,k_v,\,k_s,\,k_\theta)
\times \nc^{1-k_\theta}\times \partial^{\,k}\times \phi^n\times v^{k_v}\times  
s^{k_s}\times\theta^{k_\theta}\nonumber
\fs\eea
The first sum accounts for the
contributions generated by graphs that do not contain quark lines. It 
represents the
effective Lagrangian of gluodynamics, which we discussed in 
section \ref{gluodynamics}. The integers $k$ and $k_\theta$ count the number
of derivatives and the number of times the external field $\theta(x)$ occurs, 
respectively. In eq.~(\ref{derivative expansion}), that Lagrangian
is written in the form
$-\nc^2e_0(\vartheta)+\nc^2\partial^\mu\vartheta\partial_\mu\vartheta
e_1(\vartheta) +\ldots$, with $\vartheta=\theta/\nc$. 
As we are now expanding in powers of $1/\nc$, the vacuum energy density 
$\nc^2 e_0(\vartheta)$
is replaced by the series
$\nc^2 e_0(0)+\frac{1}{2}\,\theta^2 e_0''(0)+\ldots$ and likewise for the
other coefficients. The comparison shows that 
the coupling constant $e(k,\,k_\theta)$ is at most of order $1$.
On account of parity, $e(k,\,k_\theta)$
vanishes unless $k_\theta$ is even. Disregarding the term 
$e(0,0)=-\nc^2e_0(0)$,
which merely contributes to the cosmological constant, the sum 
only starts at $k_\theta=2$. 

The second part of the effective Lagrangian arises from graphs containing
at least one quark loop. The integers $k$ and $n$ count the number of 
derivatives and meson fields, $k_v$ is the number of external 
vector and axial fields, 
$k_s$ counts the scalar and pseudoscalar ones, while $k_\theta$ is the number
of times the field $\theta(x)$ enters. As discussed above in connection with
the scattering matrix, the translation of the 
counting rule for the matrix elements into one for the vertices
of the effective Lagrangian involves ambiguities related to the freedom
in the choice of the dynamical variables. In the presence of external fields,
that freedom becomes even richer, because the variables $\phi^a(x)$ may be 
subject to transformations that depend on these fields. It suffices to
observe, however, that the one particle irreducible matrix elements
are polynomials in the momenta, so that their Fourier transforms represent
a collection of delta-functions and derivatives thereof. We may simply
multiply this object with the relevant external fields and add a factor of
$F\phi^a(x)$ for each one of the on-shell mesons. Integrating all but
one of the coordinates over space, we obtain a specific representation for 
a term in the effective Lagrangian, for which the relevant tree graph does
reproduce the matrix element in question. The effective coupling
constants $ g(k,\,n,\,k_v,\,k_s,\,k_\theta)$  occurring therein are the 
coefficients of the polynomial that
describes the matrix element. Their order in the $1/\nc$
expansion  is determined by the counting rule (\ref{countGn}), which thus
implies that, if the effective Lagrangian is constructed in this manner, the
coupling constants $ g(k,\,n,\,k_v,\,k_s,\,k_\theta)$ are at most of order 1. 
The Lagrangian in eq.~(\ref{LN}) is what remains if all external fields are 
turned off: $g(k,n)\equiv\nc \,g(k,n,0,0,0)$. 

\subsubsection*{Chiral symmetry}
We now consider the simultaneous expansion in powers of momenta (or
derivatives) 
and $1/\nc$, introduced in section \ref{simultaneous expansion}. 
The above symbolic expression for the effective Lagrangian explicitly
displays the number of derivatives, but only indicates the leading power of
$\nc$: The expansion of the effective coupling constants in powers of $1/\nc$
starts at $O(1)$, but also contains terms of nonleading order. 
In the following, we stick to this abbreviated notation.
With the assignments specified in eqs.~(\ref{delta1}), (\ref{delta2}),
the general vertex represents a term of order
\begin{eqnarray*} \hspace{0.6em} \nc^{1-k_\theta}\times \partial^{\,k}\times
  \phi^n\times 
v^{k_v}\times   
s^{k_s}\times\theta^{k_\theta}=O(\delta^\kappa)\co\hspace{1.5em}
\kappa=\mbox{$\frac{1}{2}$}(k+k_v)
+k_s+k_\theta-1\fs\nonumber \end{eqnarray*}
Lorentz invariance implies that $k+k_v$ is even, so
that only integer powers of $\delta$ occur.
Ordering the Lagrangian in this manner, it takes the form
\bea {\cal L}_{\eff}={\cal L}^{(-1)}+{\cal  L}^{(0)}+{\cal L}^{(1)} +{\cal
  L}^{(2)}+    \ldots\;\co\nonumber\eea
where the term ${\cal L}^{(n)}$ collects all contributions of
$O(\delta^{n})$. Actually, as shown below, chiral symmetry implies that the
first term vanishes. The
expansion only begins with the term ${\cal L}^{(0)}$, which collects
the contributions of $O(1)$. 
 Note that only a finite number
of derivatives and external fields can occur at any finite order in $\delta$,
so that the relevant Lagrangian only involves a finite number of coupling 
constants, like in the standard framework.
In particular, at $O(\delta^n)$, the dependence on the vacuum angle 
is a polynomial of degree $n+2$. This feature reflects the fact that, in
the large $\nc$ limit, the $\theta$ dependence is suppressed, both in
gluodynamics and in massless QCD.
 
As mentioned in section \ref{effective action}, the Ward identities of chiral
symmetry are equivalent to the statement that the effective action is
invariant under the U(3)$_{\indR}\times$U(3)$_{\indL}$ 
gauge transformation of the external fields specified in eq.~(\ref{trafo}).
Note that the transformation law only relates quantities of the same order
in $\delta$. The effective Lagrangian can therefore be gauge invariant 
only if this is the case separately for each one of the terms 
${\cal L}^{(n)}$. In particular, the contributions of order $\delta^{-1}$ must 
altogether represent a gauge invariant expression. Now, these arise from
$k=k_v=k_s=k_\theta=0$ and thus only depend on the meson field.
We may think of this part of the effective Lagrangian as being a function
of the field $U(x)$, which, moreover, does not involve derivatives thereof, 
${\cal L}^{(-1)}=f(U)$. Invariance under  U(3)$_{\indR}\times$U(3)$_{\indL}$ 
implies that this function obeys $f(V_{\indR}\,U\,V_{\indL}^\dagger)=f(U)$.
With $V_{\indL}=U$, $V_{\indR}={\bf 1}$, this leads to 
$f(U)=\mbox{const.}$ Hence ${\cal L}^{(-1)}$ only contributes to the
cosmological constant and may be discarded.

It is convenient to replace the vacuum angle by
$\tpsi= \psi+\theta$  
and to use the fields $U,v,a,s,p,\tpsi$ and 
  their derivatives as 
independent variables -- the Lagrangian ${\cal L}^{(n)}$
is a gauge invariant function thereof. Moreover, this function is a 
polynomial in all variables except $U$. Note that only the fields
at one and the same point of space-time enter, so that it is legitimate to
treat the derivatives
$\partial_\mu U,\,\partial_\mu\partial_\nu U\,\ldots$ as independent from $U$.
Since $\tpsi$ by itself is invariant under 
U(3)$_\indR\times$U(3)$_\indL$, gauge invariance does not restrict the
dependence on this variable and its derivatives. 

At leading order, the above counting rule permits five independent
invariants\footnote{Note that the term
$\langle U^\dagger D^\mu U\rangle \langle U^\dagger D_\mu U\rangle$ 
involves two flavour traces and thus only occurs at first nonleading order
of the expansion.}: $\langle U^\dagger D^\mu D_\mu U\rangle$, 
$\langle D_\mu U^\dagger D^\mu U\rangle$, $\langle (s+ip)\, U^\dagger
\rangle$, $\langle (s-ip)\,U\rangle$, $\tpsi^2$.  The first term
differs from the second only by a total derivative and can thus be discarded.
For the Lagrangian to be real, the coefficients of $\langle (s+ip)\, U^\dagger
\rangle$ and $\langle (s-ip)\,U\rangle$ must be complex conjugates of one
another and parity then implies that they are real. Hence the leading
order Lagrangian contains three independent coupling constants. 
The explicit expression is given in eq.~(\ref{L0}). 

It is
straightforward to generalize this procedure to find the expressions for the
higher order Lagrangians. At order $\delta$, for instance, the relevant terms
are those 
of order $p^2$ with two flavour traces and those of order 
$p^4$ with one trace. Using the equations of motion associated with ${\cal
  L}^{(0)}$, the Lagrangian of order $\delta$ may be brought
to the form given in eq.~(\ref{L1}). 

\setcounter{equation}{0}
\section{Full second order SU(3) Lagrangian} 
\label{matching2}

In ref.~\cite{GL SU(3)}, the effective Lagrangian relevant for the low energy
analysis of matrix elements involving the winding number density
$\omega(x)$ or the singlet vector and axial currents $V^0_\mu(x),A^0_\mu(x)$ 
was given only at leading order. In the present appendix we briefly discuss 
the extension needed to study these quantities to first nonleading order of the
$p$-expansion. In the language of the effective theory, the relevant
extension is obtained by introducing three additional external fields,
$\theta(x)$, $\langle v_\mu(x)\rangle$, $\langle a_\mu(x)\rangle$ and 
replacing the symmetry group SU(3)$_{\indR}\times$SU(3)$_{\indL}$ by 
U(3)$_{\indR}\times$U(3)$_{\indL}$. 

The leading order SU(3) Lagrangian is given in eq.~(\ref{LSU3}).
At order $p^4$, the terms listed in ref.~\cite{GL SU(3)} can be taken over 
as they are, simply replacing the variables $U$, $a_\mu$ and $\nabla_\mu U$ 
with the quantities $\Ubar$, $\bar{a}_\mu$ and $D_\mu\Ubar$, 
defined in section \ref{dep theta}:
\bea {\cal L}^{\SU}_A\al=\al \LSU_1\langle D_\mu \Ubar^\dagger 
D^\mu \Ubar \rangle^2 +\LSU_2 \langle D_\mu \Ubar^\dagger D_\nu \Ubar
\rangle \langle D^\mu \Ubar^\dagger D^\nu \Ubar
\rangle\no\al\al +\LSU_3 \langle D_\mu \Ubar^\dagger  D^\mu \Ubar
D_\nu \Ubar^\dagger D^\nu \Ubar\rangle+
\LSU_4 \langle D_\mu \Ubar^\dagger D^\mu \Ubar\rangle
\langle\Ubar^\dagger\chi  +  \chi^\dagger\Ubar\rangle\no\al\al+
\LSU_5 \langle D_\mu \Ubar^\dagger D^\mu \Ubar (
 \Ubar^\dagger \chi+  \chi^\dagger\Ubar)\rangle +\LSU_6\langle\Ubar^\dagger
 \chi 
 +\chi^\dagger \Ubar\rangle^2\no\al\al +\LSU_7\langle 
\Ubar^\dagger \chi- \chi^\dagger \Ubar\rangle^2 +
\LSU_8\langle \Ubar^\dagger\chi\Ubar^\dagger\chi +\chi^\dagger \Ubar 
\chi^\dagger\Ubar\rangle
\\\al\al-i \LSU_9\langle \bar{R}_{\mu\nu} D^\mu \Ubar D^\nu \Ubar^\dagger+
\bar{L}_{\mu\nu} D^\mu \Ubar^\dagger D^\nu \Ubar\rangle+\LSU_{10} \langle
\bar{R}_{\mu\nu}\Ubar \bar {L}^{\mu\nu} \Ubar^\dagger\rangle
\no\al\al  
+\HSU_1\langle \bar{R}_{\mu\nu} \bar{R}^{\mu\nu}+
\bar{L}_{\mu\nu} \bar{L}^{\mu\nu}\rangle+\HSU_2\langle\chi^\dagger \chi\rangle
\co\nonumber\eea
where $\bar{R}_{\mu\nu}$,  $\bar{L}_{\mu\nu}$ are the field strengths
of $v_\mu\pm\bar{a}_\mu$.

In addition, chiral
symmetry, parity and charge conjugation invariance permit 11 new couplings, 
containing the covariant derivative of the vacuum angle or the field 
strengths of the singlet external
fields\footnote{In ref.~\cite{Hansen Leutwyler}, only the coupling constants
relevant for the two-point-functions were considered. For a corresponding list
of U(3) invariants, see \cite{Herrera 1}.}:
   \begin{eqnarray}
   {\cal L}^{\SU}_B \al=\al - i \LSU_{11}\, 
D_\mu \theta \br
   \bar{U}^\dega D^\mu \bar{U} D_\nu 
   \bar{U}^\dega 
   D^\nu \bar{U} \ke + 
   \LSU_{12} D_\mu \theta D^\mu \theta \br D_\nu
   \bar{U}^\dega 
   D^\nu \bar{U} \ke \nonumber \\
   \al\al + \LSU_{13}\, D_\mu \theta D_\nu  \theta \br D^\mu \bar{U}^\dega
   D^\nu \bar{U}\ke+ \LSU_{14}  D_\mu \theta D^\mu \theta \br \bar{U}^\dega 
\chi 
   +\chi^\dega 
   \bar{U} \ke \nonumber \\
   \al\al 
   - i \LSU_{15}\,  D_\mu \theta \br D^\mu \bar{U}^\dega \chi-  D^\mu
   \bar{U} 
   \chi^\dega \ke 
    +i \LSU_{16} \partial_\mu D^\mu \theta \, \br \bar{U}^\dega \chi
   - \chi^\dega \bar{U} 
   \ke \non \\ \al\al+i \LSU_{17} \epsilon^{\mu\nu\rho\sigma} 
   D_\mu \theta \br \bar{R}_{\nu\rho} 
   D_\sigma 
   \bar{U} \bar{U}^\dega -  \bar{L}_{\nu\rho} 
   \bar{U}^\dega D_\sigma \bar{U} \ke \\ \al\al +\HSU_3  \br
   R_{\mu\nu}+L_{\mu\nu}\ke\br R^{\mu\nu}
+L^{\mu\nu}\ke+\HSU_4  \br
   R_{\mu\nu}-L_{\mu\nu}\ke\br R^{\mu\nu}-L^{\mu\nu}\ke  \no \al\al+ 
   \HSU_5 ( D_\mu \theta D^\mu \theta )^2 + \HSU_6 ( \partial_\mu
   D^\mu 
   \theta )^2  \fs \non
   \end{eqnarray}
Finally, the anomalies of the underlying theory require an extra term that
is not gauge invariant, but does not involve any free constants. In the
notation introduced in eq.~(\ref{SWZW}), this term is given by
\bea \al\al 
\int\!\! dx{\cal L}^{\SU}_{\mbox{\tiny WZW}}\equiv 
S_{\mbox{\tiny WZW}}\{\Ubar,v,\bar{a}\}\fs
\eea
The full effective Lagrangian of order $p^4$ reads
\bea {\cal L}^{\SU}_{p^4}= 
{\cal L}^{\SU}_A+
{\cal L}^{\SU}_B+{\cal L}^{\SU}_{\mbox{\tiny WZW}} \fs\eea

Formally, the extension of the low energy analysis to the matrix 
elements of the operators $\omega(x)$, $A^0_\mu(x)$, $V^0_\mu(x)$  
thus nearly doubles the number of effective coupling constants.
Four of these, however, represent 
contact terms and a fifth
only concerns matrix elements of unnatural parity (note that a coupling
of this type does not occur in the Lagrangian of ref.~\cite{GL SU(3)} --
$\LSU_{17}$ only matters for matrix elements that involve
the winding number density or the singlet axial current). 

The inclusion of the scale dependent field $\br a_\mu \ke$ implies
that some of the coupling constants must be renormalized 
also in this version of the theory. The renormalization procedure does,
however, not entangle coupling constants occurring at different orders in the
expansion. In fact, in the above basis, the renormalization is homogeneous: 
In contrast to  
$a_\mu$, the field $\bar{a}_\mu$ is renormalization group invariant. 
The constants $\LSU_1,\,\ldots,\,\LSU_{10}$ are therefore scale independent, 
while the remaining couplings pick up a multiplicative
renormalization that compensates the one of the external fields,
\bea \al\al (D_\mu\theta)^{ren}= \za^{-1} D_\mu\theta\co\no 
\al\al\langle R_{\mu\nu}^{ren}+ L_{\mu\nu}^{ren}\rangle=
\langle R_{\mu\nu}+ L_{\mu\nu}\rangle\co\hspace{0.5em}
\langle R_{\mu\nu}^{ren}- L_{\mu\nu}^{ren}\rangle=
\za^{-1}\langle R_{\mu\nu}- L_{\mu\nu}\rangle\fs\nonumber\eea 

The renormalizations needed to absorb the infinities 
generated by the one loop graphs of ${\cal
  L}_{p^2}^{\SU}$ may be worked out as follows. The change of variables
$\hat{U}=e^{\frac{i}{3} \theta} \Ubar$ takes that Lagrangian into
\begin{eqnarray}
   {\cal L}_{p^2}^{\SU} \al = \al \tfrac{1}{4} F^2 \br 
D_\mu \hat{U}^\dega   D^\mu \hat{U} +  \hat{U}^\dega \chitheta +
\chitheta^\dagger \hat{U} \ke  +\tfrac{1}{12} \hat{H}_0 D_\mu \theta 
   D^\mu \theta \co  \label{L2su32}\no 
  \det\,\hat{U}\al=\al 1\co\hspace{1em}D_\mu\hat{U}=
\partial_\mu\hat{U}-i\,(\hat{v}_\mu+\hat{a}_\mu)\hat{U}+
i\,\hat{U}(\hat{v}_\mu-\hat{a}_\mu)\co\nonumber\end{eqnarray}
where $\hat{v}_\mu$ and $\hat{a}_\mu$ are the traceless parts of the
external fields,  $\chitheta = e^{\frac{i}{3}\theta} 
\chi$ and $\hat{H}_0 = \HSU_0 +F^2$.
In this form, the angle $\theta$ and the singlet
axial field exclusively occur in $\chitheta$ and $D_\mu
\theta$, so that the one loop calculation of ref.~\cite{GL SU(3)} can be taken
over as it is, simply replacing $\chi$ by $\chitheta$. In particular,
that calculation shows that divergences proportional to the new terms 
do not occur, so that the corresponding renormalization coefficients
$\Gamma_{11},\ldots,\,\Gamma_{17}$ and $\Delta_3,\ldots,\,\Delta_6$
all vanish.

The matching relations for the standard part ${\cal
  L}^{\SU}_A$ 
of the SU(3) Lagrangian
are given in section \ref{comparison}. These specify the
leading terms in the $1/\nc$ expansion of the coupling constants
  $\LSU_1$, $\ldots$ , $\LSU_{10}$. 
For those in the part involving the singlet external fields, the
analogous relations read\footnote{Note that the 
independence with respect to
  changes in the chiral renormalization scale is not manifest in these
  equations -- the dependence of the coupling constants on the
  scale $\mu$ is an effect of 
  order $\nc^0$ and thus beyond the given accuracy.}
\begin{eqnarray}
  \al\al \LSU_{11} = -4 (L_2 + \tfrac{1}{3} L_3) +O(1) 
   \non \\
  \al\al \LSU_{12} = \tfrac{2}{3}  (L_1 + \tfrac{1}{2} L_2 + \tfrac{1}{3} L_3)
   +O(1)   
   \non \\
\al\al   \LSU_{13} = \tfrac{4}{3} (L_2 + \tfrac{1}{3} L_3) +O(1) 
   \non \\ 
\al\al   \LSU_{14} = \tfrac{1}{3}(L_4 + 3  L_5 + L_{18} )+O(1) 
   \non \\  
\al\al   \LSU_{15} = \tfrac{1}{3}(2  L_5 + 3  L_{18} )+O(1) 
   \non \\  
\al\al   \LSU_{16} =  -  F^4 (1+\Lambda_1)(1+\Lambda_2) \,(72\,\tau)^{-1}
   +O(1) \label{Kmatch}
    \\
\al\al   \LSU_{17} = {\nc}({288\, \pi^2})^{-1}+\tfrac{1}{2} \tilde{L}_4
   +O(\nc^{-1})  
   \non \\  
\al\al   \HSU_3 =  O(1) 
   \non \\ 
\al\al   \HSU_4 =   \tfrac{1}{6} (H_{1}-\tfrac{1}{2}L_{10} )+O(1) 
   \non \\ 
\al\al   \HSU_5 = \tfrac{1}{9}  (L_1 + \tfrac{1}{2} L_2 + \tfrac{1}{3} L_3)
   +O(1) 
   \non \\
\al\al   \HSU_6 =  F^4 (1+\Lambda_1)^2\,(72\,\tau)^{-1} +O(1)
   \fs \nonumber
   \end{eqnarray}
The coupling constants $\LSU_{16}$ and $\HSU_6$ receive a
contribution from $\eta'$-exchange, similar to the one in $\LSU_{7}$
(section \ref{comparison}). The leading contribution to $\LSU_{17}$ stems
from the Wess-Zumino-Witten term of the extended theory (appendix
\ref{decomposition}).

Note that, in the SU(3) framework, the Kaplan-Manohar
transformation (\ref{KM}) takes the mass term of the effective
Lagrangian into
\bea \langle \bar{U}^\dagger\chi \rangle\rightarrow
\langle\bar{U}^\dagger \chi
\rangle + \frac{\lambda}{4 B} \left\{ \br \chi^\dagger \bar{U} \ke^2 -\langle
  \chi^\dagger  \bar{U} \chi^\dagger
  \bar{U}  
\rangle \right\} \co\nonumber\eea 
even if the vacuum angle does not vanish: The factor
$e^{-i\theta}$ is absorbed in the field $\Ubar$, 
on account of $\det \Ubar =e^{-i\theta}$. Hence the transformation of the quark
mass matrix in eq.~(\ref{KM}) is equivalent to a change of the effective 
coupling constants occurring in ${\cal L}^{\SU}_{p^4}$, 
also for $\theta\neq 0$.
 
\setcounter{equation}{0}
\section{Renormalization of the WZW term}
\label{decomposition}
We first observe that the field
$\bar{U}$ introduced in eq.~(\ref{Ubar})
is renormalization group invariant. Moreover, $U$ and $\bar{U}$ transform in
the same way under chiral rotations, 
because the factor that makes the difference,
$e^{\frac{i}{3}\tpsi}$, is invariant.  
This immediately implies that the anomalies
of the functional $S_{\mbox{\tiny{WZW}}}$ remain the same if $U$ is replaced
by $\bar{U}$. Indeed, performing the above change of
variables in eq.~(\ref{SWZW}), we obtain a gauge invariant result
for the difference:
\bea \al\al S_{\mbox{\tiny{WZW}}}\{U,v,a\}=
S_{\mbox{\tiny{WZW}}}\{\bar{U},v,a\}
+\int\!\! A\co\nonumber\\
\al\al A= - \frac{\nc}{144\pi^2}\, \tpsi\,\langle 
i F_{r}\, DU DU^\dagger
+F_{r}\,U F_{l}\,U^\dagger +2 F_{r}^2 +\RL\,\rangle
\co\nonumber\\
\al\al F_{r}=dr-i r^2\co\hspace{2em}
F_{l}=dl-i l^2\fs\nonumber\eea
with $DU=dU-irU+iUl$. The terms occurring here are of the
same form as those entering the U(3)$_{\indR}\times$U(3)$_{\indL}$ invariant 
part of the
effective Lagrangian. 

Next, we decompose the axial field according to eq.~(\ref{ab}). The
corresponding decomposition of the right- and lefthanded gauge fields
into a renormalization group invariant part $\bar{r},\bar{l}$ and
a remainder reads $r=\bar{r}+\frac{1}{6}D\theta$,
$l=\bar{l}-\frac{1}{6}D\theta$. Using the identity
$\langle  d\bar{U}\bar{U}^\dagger\rangle=-id\theta$, which follows 
from $\det\bar{U}=e^{-i\theta}$, we then obtain
\bea \al\al S_{\mbox{\tiny{WZW}}}\{U,v,a\}=
S_{\mbox{\tiny{WZW}}}\{\Ubar,v,\bar{a}\}+\int (A+B+ P_1+P_2)\co\no
\al\al B= 
\frac{ \nc}{144\pi^2}\,i D\theta\,\langle \bar{F}_{r}\,D\Ubar 
\Ubar^\dagger
- \bar{F}_{l}\,\Ubar^\dagger D\Ubar
\rangle \co\no
\al\al\bar{F}_{r}=d\bar{r}-i\bar{r}^2\co\hspace{2em}
\bar{F}_{l}=d\bar{l}-i \bar{l}^{\,2}\fs\nonumber\eea
Note that terms proportional to 
$D\theta\,\langle (D\Ubar\Ubar^\dagger)^3\rangle$
cancel out on account of charge conjugation invariance.
The term $S_{\mbox{\tiny{WZW}}}\{\bar{U},v,\bar{a}\}$ is 
renormalization group invariant. Under chiral rotations, it transforms
with
\bea \delta S_{\mbox{\tiny{WZW}}}\{\bar{U},v,\bar{a}\} =
\int\!\langle(\alpha_{\indL}-\alpha_{\indR})\Omega_0\rangle\fs\nonumber\eea
The one with $ B$ is gauge invariant, but transforms with
$\za^{-1}$ under the renormalization group. 
The calculation automatically yields the contact
terms $P_1$, $P_2$ introduced in section \ref{unnatural}, 
which account for the difference between $\Omega$ and $\Omega_0$ -- these
transform in a nontrivial manner, both under the renormalization group 
and under chiral rotations.
Finally, $A$ may also be sorted out according to the behaviour under
the renormalization group:
\bea  A\al=\al 
- \frac{\nc}{144\pi^2}\,\tpsi\,\langle 
i \bar{F}_{r}\, D\Ubar  D\Ubar^\dagger
+\bar{F}_{r}\,\Ubar \bar{F}_{l}\,\Ubar^\dagger +2
\bar{F}_{r}^2+\RL\,\rangle\no
\al\al-\frac{\nc}{216\pi^2}\,\tpsi\,\langle da\rangle\, \langle da \rangle
\fs\nonumber\eea
This completes the decomposition of $S_{\mbox{\tiny{WZW}}}\{U,v,a\}$.
The renormalization group invariant part is the Wess-Zumino-Witten term
relevant for the effective theory at fixed $\nc$:
\bea \int\!\!dx\,{\cal L}_{\mbox{\tiny WZW}}^{\SU}\equiv
S_{\mbox{\tiny{WZW}}}\{\Ubar,v,\bar{a}\}\fs\eea
The extension to the degrees of freedom carried by the $\eta'$
contains the following additional contributions, which are gauge invariant:
\bea{\cal L}_{\mbox{\tiny WZW}}\al=\al{\cal L}_{\mbox{\tiny WZW}}^{\SU}
-\frac{\nc\,\epsilon^{\mu\nu\rho\sigma}}{288\pi^2}\left\{\tpsi\,
\langle i\bar{R}_{\mu\nu}D_\rho\Ubar D_\sigma \Ubar^\dagger+i\bar{L}_{\mu\nu}
D_\rho\Ubar^\dagger D_\sigma\Ubar
+\bar{R}_{\mu\nu}\Ubar\bar{L}_{\rho\sigma}
\Ubar^\dagger\rangle\right.\no\al\al\hspace{7em} +
\tpsi\, \langle \bar{R}_{\mu\nu}\bar{R}_{\rho\sigma}+
\bar{L}_{\mu\nu}\bar{L}_{\rho\sigma}\rangle+\mbox{$\frac{1}{12 }$}\,
\tpsi\, \langle R_{\mu\nu}-L_{\mu\nu}\rangle\langle
R_{\rho\sigma}-L_{\rho\sigma}\rangle \no\al\al\hspace{7em}
 -i D_\mu\theta \,\langle 
\bar{R}_{\nu\rho}
D_\sigma \Ubar \Ubar^\dagger -\bar{L}_{\nu\rho} D_\sigma\Ubar^\dagger
\Ubar\rangle \left.\right\}\fs\eea

\end{document}